\documentclass[sn-mathphys]{sn-jnl}% Math and Physical Sciences Reference Style
\newcommand{\oph}{RS~Oph} 
\newcommand{\fermi}{\textit{Fermi}-LAT} 
\usepackage{color}
\usepackage{amssymb}
 \usepackage[T1]{fontenc} % needed for \dj
\newcommand{\aj}{Astron. J.}   % Astronomical Journal
\newcommand{\apj}{Astrophys. J.}   % Astrophysical Journal
\newcommand{\apjl}{Astrophys. J. Lett.}   % Astrophysical Journal, Letters
\newcommand{\apjs}{Astrophys. J. Suppl. Ser.}   % Astrophysical Journal, Supplement
\newcommand{\aap}{Astron. Astrophys.}   % Astronomy and Astrophysics
\newcommand{\mnras}{Mon. Not. R. Astron. Soc.}   % Monthly Notices of the RAS
\newcommand{\nat}{Nature} % Nature
\newcommand{\na}{New Astron.}   % New Astronomy
\newcommand{\prd}{Phys. Rev. D}   % Physical Review D
\jyear{2021}%

\theoremstyle{thmstyleone}%
\theoremstyle{thmstyletwo}%

\theoremstyle{thmstylethree}%

\begin{document}

\title{Proton acceleration in thermonuclear nova explosions revealed by gamma rays}

%%=============================================================%%
%% Prefix	-> \pfx{Dr}
%% GivenName	-> \fnm{Joergen W.}
%% Particle	-> \spfx{van der} -> surname prefix
%% FamilyName	-> \sur{Ploeg}
%% Suffix	-> \sfx{IV}
%% NatureName	-> \tanm{Poet Laureate} -> Title after name
%% Degrees	-> \dgr{MSc, PhD}
%% \author*[1,2]{\pfx{Dr} \fnm{Joergen W.} \spfx{van der} \sur{Ploeg} \sfx{IV} \tanm{Poet Laureate} 
%%                 \dgr{MSc, PhD}}\email{iauthor@gmail.com}
%%=============================================================%%

%\author{MAGIC~Collaboration,$^{1-52, \ast,}$}
%\equalcont{\small{A list of authors and their affiliations appears at the end of the paper}\\
%\small{$^\ast$corresponding authors(contact.magic@mpp.mpg.de): J.~Sitarek, R.~L\'opez-Coto, D.~Green, A.~L\'opez-Oramas}}
\author{\fnm{V.~A.} \sur{Acciari}$^{1}$} 
\author{\fnm{S.} \sur{Ansoldi}$^{2,42}$} 
\author{\fnm{L.~A.} \sur{Antonelli}$^{3}$} 
\author{\fnm{A.} \sur{Arbet Engels}$^{15}$} % ETH(4) ==> MPI (15)
\author{\fnm{M.} \sur{Artero}$^{5}$} 
\author{\fnm{K.} \sur{Asano}$^{6}$} 
%\\
\author{\fnm{D.} \sur{Baack}$^{7}$} 
\author{\fnm{A.} \sur{Babi\'c}$^{8}$} 
\author{\fnm{A.} \sur{Baquero}$^{9}$} 
\author{\fnm{U.} \sur{Barres de Almeida}$^{10}$} 
% ...

\affil[1]{\orgname{ Instituto de Astrof\'isica de Canarias and Dpto. de  Astrof\'isica, Universidad de La Laguna, } \orgaddress{E-38200, La Laguna, Tenerife, Spain}}
\affil[2]{\orgname{ Universit\`a di Udine and INFN Trieste, } \orgaddress{I-33100 Udine, Italy}}
\affil[3]{\orgname{ National Institute for Astrophysics (INAF), } \orgaddress{I-00136 Rome, Italy}}
% ...
\affil[5]{\orgname{ Institut de F\'isica d'Altes Energies (IFAE), The Barcelona Institute of Science and Technology (BIST), } \orgaddress{E-08193 Bellaterra (Barcelona), Spain}}
\affil[6]{\orgname{ Japanese MAGIC Group: Institute for Cosmic Ray Research (ICRR), The University of Tokyo, } \orgaddress{Kashiwa, 277-8582 Chiba, Japan}}
\affil[7]{\orgname{ Technische Universit\"at Dortmund, } \orgaddress{D-44221 Dortmund, Germany}}
\affil[8]{\orgname{ Croatian MAGIC Group: University of Zagreb, Faculty of Electrical Engineering and Computing (FER), } \orgaddress{10000 Zagreb, Croatia}}
\affil[9]{\orgname{ IPARCOS Institute and EMFTEL Department, Universidad Complutense de Madrid, } \orgaddress{E-28040 Madrid, Spain}}
\affil[10]{\orgname{ Centro Brasileiro de Pesquisas F\'isicas (CBPF), } \orgaddress{22290-180 URCA, Rio de Janeiro (RJ), Brazil}}
% ...
\affil[15]{\orgname{ Max-Planck-Institut f\"ur Physik, } \orgaddress{D-80805 M\"unchen, Germany}}
% ...
\affil[42]{\orgname{ also at International Center for Relativistic Astrophysics (ICRA), } \orgaddress{Rome, Italy}}
% ...
\affil[58]{\orgname{ANS Collaboration}, \orgaddress{c/o Astronomical Observatory, I-36012 Asiago (VI), \country{Italy}}}

\maketitle

\textbf{
Classical novae are cataclysmic binary star systems in which the matter of a companion star is accreted on a white dwarf (WD) \cite{2008clno.book.....B, doi:10.1146/annurev-astro-112420-114502}.
Accumulation of hydrogen in a layer eventually causes a thermonuclear explosion on the surface of the WD \cite{2004ApJ...600..390T}, brightening the WD to $\sim10^5$ solar luminosities and triggering ejection of the accumulated matter. 
  They provide extreme conditions required to accelerate particles, electrons or protons, to high energies. 
  Here we present the detection of gamma rays by the MAGIC telescopes from the 2021 outburst of RS Ophiuchi (\oph{}), a recurrent nova with a red giant (RG) companion, that allowed us, for the first time, to accurately characterize the emission from a nova in the 60\,GeV to 250\,GeV energy range.
  The theoretical interpretation of the combined \fermi{} and MAGIC data suggests that protons are accelerated to hundreds of GeV in the nova shock. 
  Such protons should create bubbles of enhanced Cosmic Ray density, on the order of 10\,pc, from the recurrent novae.
%   Such protons should create bubbles of enhanced Cosmic Ray density up to the order of 10\,pc from the recurrent novae.
}

%\section*{Introduction}

%Classical novae are cataclysmic binary star systems in which the matter of a companion star is accreted on a white dwarf (WD) \cite{2008clno.book.....B, doi:10.1146/annurev-astro-112420-114502}.
%Accumulation of the matter in a layer eventually causes a thermonuclear explosion on the surface of the WD \cite{2004ApJ...600..390T}, brightening the WD to $\sim10^5$ solar luminosities and triggering ejection of the accumulated matter. 
%If the companion star of the WD is a red giant (RG), such a system can form a symbiotic nova \cite{2012BaltA..21....5M}.
A symbiotic nova can be formed when the companion star of the WD is a RG. %red giant 
%(RG) 
\cite{2012BaltA..21....5M}. 
The ejecta of symbiotic novae expand within the dense wind of the RG companion. 
Novae outbursts usually last from weeks to months.
While they are expected to repeat hundreds of times \cite{1978ApJ...219..595F}, the interval between subsequent events can be even hundreds of thousand years \cite{2010ApJS..187..275S}.
However, a subclass of objects called Recurrent Novae (RNe) allows one to observe such repeated outbursts over a human lifespan \cite{1987ApJ...314..653W}. 
In our Galaxy, ten such objects are known in which the repetition of bursts has been seen within a century \cite{2010ApJS..187..275S}.
According to \cite{2008ASPC..401...42M} for the symbiotic nova to become recurrent, its WD must be massive ($\geq$1.1 M$_{\odot}$). 

Novae have been deeply studied in the optical and X-ray ranges for decades \cite{1997ApJ...491..312C, 2006Natur.442..276S, 2006ApJ...652..629B, 2007ApJ...665.1334N,2008ApJ...673.1067N, 2010ApJS..187..275S}, but only recently they have been shown as emitters of high-energy gamma-ray radiation: first in the case of symbiotic novae \cite{2010Sci...329..817A} and soon after with classical novae \cite{2014Sci...345..554A}.
Though this clearly indicates that charged particles are accelerated to high energies in novae, their nature and radiation mechanism are not yet clear. 
In order to understand the acceleration mechanism of high-energy particles, it is crucial to measure the maximum energies of the emitted radiation. Until recently, all spectra of gamma-ray novae have been  measured only up to 6 -- 10 \,GeV range \cite{2014Sci...345..554A} with no hint of emission at higher energies \cite{2012ApJ...754...77A,2015A&A...582A..67A}.

%\subsection*{The 2021 flare in the recurrent symbiotic nova \oph{}}
\oph{} is a recurrent symbiotic nova with average time between major outbursts of 14.7 years \cite{2010ApJS..187..275S}. 
The latest outburst, in August 2021, was promptly reported in optical \cite{KGeary2021} and high-energy (HE, $100\,\mathrm{MeV}<E<10\,\mathrm{GeV}$) gamma rays by \fermi{} \cite{2021ATel14834....1C}. 
The optical emission showed similar behaviour to the 2006 outburst (see Extended Data Figure EDF~1.%\ref{figs:photometry})
Following these alerts, MAGIC began observations of \oph{} as part of its nova follow-up program \cite{2015A&A...582A..67A}, on August 09, 2021 at 22:27 UT, i.e., about 1 day after the first optical and GeV detections. 
In parallel, the H.E.S.S. collaboration announced very-high-energy (VHE, $\gtrsim 100$\,GeV) gamma rays from \oph{} \cite{2021ATel14844....1W}.
The MAGIC observations reveal VHE emission contemporaneous to the \fermi{} and optical maxima, and a decrease below the VHE detection limit two weeks later (see Fig.~1).
Details of the analysis can be found in Methods section~\ref{sec:magic}. 
The first four days of MAGIC observations (August 09-12) yield a VHE signal with a significance of 13.2\,$\sigma$ (see EDF~2), %\ref{figs:theta2}
spanning from 60 GeV to 250\,GeV, well fitted by a single power-law ($\chi^2/N_{dof} = 5.9/5$). 
 
Daily spectra are reconstructed (see EDF~3, %\ref{figs:sed_daily}
Method sections \ref{sec:magic} and Supplementary section~H%\ref{sect:days}
) allowing us to track the evolution of the outburst.

The contemporaneous gamma-ray spectrum measured by \fermi{} and MAGIC can be described as a single, smooth component spanning from 50\,MeV to 250\,GeV.
Intriguingly, while the GeV emission subsides with a halving time scale of $\sim2.2$ days (see also Methods section~\ref{sect:fermi}), the flux measured by MAGIC over the first four days is consistent with being constant (\mbox{$\chi^2/\mathrm{N_{dof}}$ = 2.9/3}), see also EDF~4%\ref{figs:fermi_magic_optical}. 
This suggests a migration of the gamma-ray emission towards higher energies, in line with an increase of the maximum energies of the parent particles. 
\oph{} is the gamma-ray nova with the highest flux and energy output to date, as shown by the comparison with the other \fermi{} detected novae presented in Supplementary section~I%\ref{sec:comparison_novae}
.
Therefore, the non-detection of previous novae at VHE range \cite{2012ApJ...754...77A,2015A&A...582A..67A} might be explained by the lack of sensitivity to dimmer eruptions, without the need to invoke any fundamental difference in the spectral energy distribution of \oph{}. 

%\subsection*{Novae as high-energy particle accelerators}

The conditions in novae are favourable for the acceleration and subsequent emission of radiation by both electrons and protons \cite{2014Sci...345..554A}.
The expanding ejecta of a nova interacting with the interstellar medium (filled also with the dense RG wind in the case of symbiotic novae) will result in the formation of a shock wave. 
Moreover, the fast wind, induced by the nuclear burning on the surface of the WD, will catch up with the ejecta, causing an additional internal shock \cite{2018A&A...612A..38M}.
Recently, a correlation between optical and gamma-ray emission has further suggested that a substantial part of the novae explosion's power goes into shocks \cite{2020NatAs...4..776A}.
%\cite{2014Sci...345..554A}.
In such shocks, energetic electrons and protons can be produced (see Fig.~2). 
Gamma-ray emission can arise from photosphere thermal radiation up-scattered to the gamma-ray energy range by relativistic electrons via inverse Compton scattering.
Alternatively, the ambient matter (nova ejecta and RG wind) can act as a target for hadronic interaction of protons or Bremsstrahlung radiation of electrons \cite{2014Sci...345..554A}.
The maximum energies of high-energy particles will depend on the efficiency of the acceleration mechanism, duration of the nova, and the cooling energy losses (see Methods section \ref{sect:maxe} and EDF~5%\ref{figs:timescales}
). 
Protons experience only mild cooling by proton-proton interactions with time scale of 
%= -E_p / \frac{dE_p}{dt}
$t_{pp} = 21 (n_p/6\times10^8\,\mathrm{cm^{-3}})^{-1}\,\mathrm{[day]}$,
where $n_p$ is the number density of the target material. 
Electrons in nova shocks suffer stronger inverse Compton energy losses with 
$t_{IC} = 4.4 \times 10^{-3} (E/300\,\mathrm{GeV})^{-1})[1+10(E/300\,\mathrm{GeV})]^{1.5}\,\mathrm{[day]}$.
%see Section \ref{sect:maxe}), 
Therefore, the production of high energy photons via leptonic mechanisms is much more demanding on the acceleration processes efficiency than for proton models. 
The simultaneous acceleration of both types of particles (but reaching different energies) has also been proposed \cite{2012PhRvD..86f3011S, 2015A&A...582A..67A}.
We estimate that Bremsstrahlung is negligible with respect to inverse Compton component for the parameters of \oph{} (see Methods section~\ref{sec:model}).

%\subsection*{Evidence for acceleration of protons to hundreds of GeV in RS Oph}

We derive the photosphere parameters using fits to the photometry measurements (see EDF~6%\ref{figs:spect}
) and shock expansion velocity from spectroscopy (see EDF~7%\ref{figs:line_profiles}
).
Based on the optical observations of \oph{} during the 2021 outburst, and the derived parameters from previous outbursts of the source, we model the gamma-ray emission with the injection of a population of relativistic electrons or protons (see Methods section~\ref{sec:model}).
We take into accout also the minor absorption of the emission in the photosphere radiation field (see EDF~8%\ref{figs:abs}
).
The \fermi{} and MAGIC measurement can be well described ($\chi^2/\mathrm{N_{dof}} = 13.1/12$, p-value = 0.36) with the proton-only model (see left panel of Fig.~3). 

The fit yields a canonical power-law spectrum with an index $\sim -2$ and an exponential cut-off, corresponding to the maximum energies achieved in the acceleration.
The day-by-day modeling shows evidence that the energy cut-off of protons increases with time (see Supplementary section~H %\ref{sect:days} 
and EDF~9%\ref{figs:ecut}
). 
This goes in line with absence of spectral signatures from cooling terms.
The associated neutrino emission is not expected to be detected by the current experiments (see Supplementary section~F%\ref{sect:neutrino}
).

In contrast, it is difficult to explain the shape of the curvature of the measured spectrum between 50\,MeV and 250\,GeV with leptonic processes. 
The leptonic model requires injection of particles that already contain a strong break (change of particles index by $3.25\pm0.28$) in the electron energy distribution (see Fig.~3, right panel). 
Since the break must already be present in the injection spectrum of particles, it cannot be explained by the cooling.
In addition, despite a more complicated particle injection model, the description of the gamma-ray emission in the electron scenario is significantly worse ($\chi^2/\mathrm{N_{dof}}$ = 27.5/11, p-value $= 3.9\times10^{-3}$ ) than in the case of protons, as can be seen in Fig.~3. 
The relative likelihood of the electron model with respect to the proton model for $\Delta$AIC$=15.3$, as defined within the Akaike information criterion framework \cite{1974ITAC...19..716A}, which is normally used for comparison of non-nested models, is $4.7\times 10^{-4}$.

Despite their intense emission of gamma rays, accelerated protons will eventually escape the nova shock carrying away most of their obtained energy. % and contribute to the CR sea
Such protons can contribute to the Galactic Cosmic Rays (CR), which are expected to be produced mainly in supernova remnants \cite{2016MNRAS.457.1786M}. 
 
The measurement of the proton spectrum required to explain the gamma-ray emission of \oph{} can be used to put estimates on novae contribution to CR. 
Using the CR energetics derived for \oph{} ($\sim4.4 \times 10^{43}$\,erg, see Methods section~\ref{sect:energy}), a rate of $50$ novae per year \cite{2017ApJ...834..196S} would lead to about 0.1\% of the CR energy contribution from supernovae, which are more rare than novae ($\sim2$ per century) but much more energetic ($\sim10^{50}$\,erg). 
Despite the small contribution to the overall CR sea, a nova would significantly increase the CR density in its close environment.
The energy density of the nova dominates over that of the average CR energy density in the Milky Way ($\sim$1.8\,eV/cm$^3$) in a region of radius $\sim$0.5 pc, of the order of the distance to the nearest star in our Galaxy. 
In the special case of recurrent novae, protons accelerated over  $10^{5}$\,yr \cite{1978MNRAS.183..515B}, assuming a recurrent rate of every 15 years, will accumulate in a $\sim 9$\,pc bubble with enhanced CR density (see  Methods section~\ref{sect:cr_sea}).

%\subsection*{Conclusions}
The detection of gamma rays reaching 250\,GeV from a recurrent symbiotic nova allowed us to obtain a deep physical insight on the population  of relativistic particles accelerated by such objects.
The modeling of the gamma-ray spectrum strongly favors the explanation of the emission via the acceleration of protons in a nova shock. 
Evidence towards the proton acceleration is based on:
(i) the inferred shape of the energy distribution of injected particles, 
(ii) the better statistical description of the gamma-ray spectral energy distribution by the proton model,
(iii) the obtained evidence of the increase of the particle maximum energies over time, consistent with lack of strong cooling. 
The protons in the nova shock undergo slow cooling, therefore they will be eventually able to escape the shock, carrying away a significant fraction of energy. 
Such protons will add to the Galactic cosmic ray budget, however primarily in the close neighborhood of novae. 

The observation of the August 2021 outburst of \oph{} introduces a new class of sources as VHE gamma-ray emitter: (recurrent symbiotic) novae.
\oph{} is a recurrent symbiotic nova, the same class of objects as V407 Cyg, the first nova detected in the GeV range by \fermi{}. 
While we now know that classical novae are also GeV emitters, it is still to be seen if the detection of \oph{} emitting in VHE gamma-ray range is due to its recurrent symbiotic nature, or just the first sign of such emission from a broader class of classical novae. 
The comparison of gamma-ray measurements in GeV and VHE gamma-ray range with previous \fermi{} novae does not reveal any peculiarity in the emission of \oph{}, except for its brightness (see Fig.~4 and EDF~10%\ref{figs:v339_and_v407_comparison}
). 
Therefore, it is likely that future, more sensitive VHE gamma-ray facilities will be able to provide an ample harvest of novae.

\backmatter

\bmhead{Supplementary information}
\ \\ 
Supplementary sections C-I \\
Extended data figures EDF 1-10 \\
Supplementary Tables 1 -- 10 \\
References (70-97) \\

%If your article has accompanying supplementary file/s please state so here. 

%Please refer to Journal-level guidance for any specific requirements.

\clearpage

%\begin{appendices}
\section*{Methods}
\appendix

\section{Observations and data analysis}

In this section we report the detailed results of the analysis of gamma-ray data with MAGIC and \fermi{}, and optical data with TJO and ANS.

\subsection{MAGIC}\label{sec:magic}

MAGIC \cite{2016APh....72...61A} is a stereoscopic system of two imaging atmospheric Cherenkov telescopes situated in the Canary island of La Palma, Spain (28.8$^\circ$N, 17.9$^\circ$W at 2225 m above sea level). Each telescope consists of a 17-m diameter mirror dish and a fast imaging camera. 
The system achieves a sensitivity of (0.92 $\pm$ 0.04)\% of the Crab Nebula flux above $210$\,GeV in $50$\,h in zenith angle range $30-45^\circ$ \cite{2016APh....72...76A}. 

MAGIC observed RS Oph in the period between August 09, 2021 to September 01, 2021 (MJD  59435.94 to  59458.97) for 34.0 h (see Supplementary Table~2%\ref{tab:magic_data}
). The  data quality selection was based on the atmospheric transmission and rates of background events. 
For this analysis we also did not include data taken under moonlight condition, as they provide much higher energy threshold values. 
After quality cuts, 21.4\,h % 10.8 + 10.6 
of the data were used for the analysis, half of which were taken during the first four days after the nova eruption. 
The source was observed at zenith angles between 36$^\circ$ and 60$^\circ$. 
The data were taken in the so-called wobble mode, pointing at four different sky positions situated 0.4$^\circ$ away from the source to evaluate the background simultaneously.

The data were analyzed using the MAGIC Analysis and Reconstruction Software, MARS \cite{Zanin2013}. 
A dedicated low-energy procedure with a special signal extraction and image cleaning, the so-called MaTaJu method, was applied (see \cite{2021A&A...647A.163M} and references therein).
Further processing of the data, including the image parameterization, the direction and energy reconstruction and gamma-hadron separation, were applied following the standard MARS analysis chain. 
The energy threshold of the analysis is $\sim$60 GeV.

% 20220216 JS removed 
%RS Oph is detected with a significance of 13.2\,$\sigma$ (see Fig.~\ref{figs:theta2}).

We fitted the spectrum obtained from the first four days of observations using a single power-law (d$N$/d$E$ = $f_0\ (E/E_0)^{-\alpha}$), 
resulting with a $\chi^2/\mathrm{N_{dof}} = 5.9/5$ goodness of fit.
The used fit also takes into account estimated energy bins without detected signal, hence the number of degrees of freedom is larger than expected from the number of points in the reconstructed spectrum.
The normalization energy of the fit (\mbox{$E_0 = 130$\,GeV}) is the decorrelation energy (i.e. normalization energy which minimizes the correlation of the fit parameters) of the four-day sample. 
The fit parameters are listed in Supplementary Table~3%\ref{tab:daily_sed}
.

In order to estimate the lower limit on the maximal true energy
of gamma rays consistent with the MAGIC data we follow the procedure of \cite{2019Natur.575..455M}. 
We perform a likelihood fit of the data with a power-law model with a sharp cut-off at a given energy $E_{cut}$. 
The $3\sigma$ (99.7\% C.L.) lower limit on the $E_{cut}$ is the value for which the increase of the $\chi^2$ of the fit is equal to 9. 
We obtain $170$\,GeV, however taking into account also the 15\% systematic uncertainty on the energy scale following \cite{2016APh....72...76A} we obtain a slightly less constraining, conservative limit of $E_{cut}>150$\,GeV.

We have also performed night-by-night spectral fits to investigate spectral variability. 
The parameters from the first two nights are consistent within errors (note however that the exposure on the first night is lower than on the remaining ones). 
A hint of hardening of the emission is seen between the second and third night.
No significant change of parameters can be seen between the third and the fourth night. 

The daily-binned light curve was calculated for an integral flux above 100 GeV. For the first four days the fit to a constant flux gives a %$\chi^2 = $xx.
$\chi^2/\mathrm{N_{dof}}$ = 2.9/3 % JS points with MaTaJu, updated 20210912
with a  value of $F_{0} = (4.41\pm0.46_{\rm stat}) \times 10^{-11}  \mathrm{\,cm^{-2}\,s^{-1}}$.

\subsection{\fermi}\label{sect:fermi}

The Large Area Telescope on-board the \textit{Fermi} Gamma-ray Space Telescope (\fermi{}), is a pair conversion telescope designed to detect gamma rays with an energy range of 0.02\,GeV to $>$ 300\,GeV \cite{2009ApJ...697.1071A}.
The \fermi{}, with its large field of view (2.4\,sr), observes the entire sky approximately every 3 hours.
%\footnote{https://fermi.gsfc.nasa.gov/ssc/observations/types/post\_anomaly/}.
Each analysis is performed with \emph{fermitools} v2.0.8
%\footnote{https://fermi.gsfc.nasa.gov/ssc/data/analysis/software/}
and \emph{Fermipy} v1.0.2 \cite{2017ICRC...35..824W} using a binned likelihood analysis, P8R3\_V3 instrument response functions (IRFs), and the catalog 4FGL-DR2 \cite{Abdollahi_2020,2020arXiv200511208B} with the standard Galactic and isotropic diffuse background to construct the model of the region of interest (ROI).  
For each analysis, the \textit{SOURCE} event class is used as this is the recommended event class for long duration observations, observations of more than a few hours. 
The \textit{SOURCE} event class can be further divided into separate event types such as \emph{PSF0}, \emph{PSF1}, \emph{PSF2}, and \emph{PSF3}, where \emph{PSF0} corresponds to events with the worst PSF and \emph{PSF3} are events with the best PSF.  

For the 1-day and 3-day time bins, the \fermi{} data-set used encompasses a total time range from MJD 59431.45 to 59461.45, an energy range from 0.1 GeV to 1000 GeV, and a $15^\circ$ ROI centered on the radio coordinates of \oph{} (R.A. = 267.555$^{\circ}$, Dec. = -6.7078$^{\circ}$).
We use event type 3, which corresponds to all events, for this analysis and select a maximum zenith angle of $>90^{\circ}$ to reduce any gamma-ray contamination from the Earth limb.  
The majority of 4FGL-DR2 sources for the one and three day time bins are not significantly detected (Test Statistic (TS) $>25$, see \cite{1996ApJ...461..396M}), apart from 4FGL J1813.4-1246 and 4FGL J1745.4-0753. 
These sources correspond to PSR J1813-1246 and TXS 1742-078, which are $8.3^\circ$ and $1.7^\circ$ away from \oph{}.
Here, TS is defined as TS = -2 $\ln$ ($\mathcal{L}_{max,0}$/$\mathcal{L}_{max,1}$), where $\mathcal{L}_{max,0}$ is the maximum likelihood of the null hypothesis and $\mathcal{L}_{max,1}$ is the maximum likelihood with the source included \cite{1996ApJ...461..396M}. The square-root of the TS is approximately equal the significance of detection, i.e. a TS = 25 is $\sim 5\,\sigma$.
TXS 1742-078 is a non-variable hard blazar and therefore could cause possible source confusion. 
Due to the proximity of 4FGL J1745.4-0753 to \oph{} and possible source confusion at the lowest energies, the value of the index of 4FGL J1745.4-0753 is locked to that of the 4FGL-DR2 catalog. 
\oph{} is included in the ROI and modeled with a Log Parabola model.
Additional spectral models were tested for a four-day period contemporaneous to MAGIC observations: a power-law (TS$ = 2168.1$) as well as a power-law  with an exponential cutoff (TS$ = 2016.4$), and the Log Parabola model (TS$ = 2226.44$) had the highest TS, and therefore we use the Log Parabola model as our spectral form for \oph{}.  
The ROI is optimized with the normalization and spectral parameters of any 4FGL-DR2 source with a number of predicted counts $<$ 1 locked to the 4FGL-DR2 values, excluding the Galactic and isotropic diffuse background.
All parameters on all unlocked 4FGL-DR2 sources within $4^\circ$ are left free to vary, and the ROI is fit using \emph{Minuit} minimizer.  
If \oph{} source model does not have a TS $>$ 9, number of predicted counts $>$ 4 or the error of the integrated flux from 0.1\,GeV to 1000\,GeV is greater than 60\% of the value, then it is not considered detected and 95\% upper limits (ULs) are calculated.  
These 1-day and 3-day light curves are presented in Supplementary Tables 4 and 5  %\ref{tab:lat_daily_flux} and \ref{tab:lat_3daily_flux} 
and in Fig.~1.
The 1-day light curve in MJD 59435.45--59444.45 %(i.e. from the peak up to the first non-detection) 
range can be well fit ($\chi^2/\mathrm{N_{dof}} = 6.5/7$) with an exponential decay with halving time of ($2.20 \pm 0.18$)\,days.  
%
% 20220216 JS removed
%The associated SEDs for the first four days are presented in Fig.~\ref{figs:sed_daily}.

The analysis of the combined first four days has a data-set which encompasses a time range MJD 59435.45 -- 59439.45 and an energy range from 0.05\,GeV to 1000\,GeV.  
Reaching down to 0.05\,GeV is necessary to help distinguish leptonic and hadronic models described in the main text and seen in Fig.~3. 
The same procedure is applied as in the 1-day and 3-day time bins, with some adjustments in the settings to allow the analysis to reach 0.05\,GeV.  
Due to the worsening of the \fermi{} PSF below 0.1\,GeV, we apply a $20^\circ$ ROI centered on \oph{}, and a more restrictive zenith angle selection of $> 80^\circ$. 
We perform a joint-likelihood analysis with two components, one in the energy range between 0.05\,GeV to 0.1\,GeV and one in the energy range from 0.1\,GeV to 1000\,GeV.  
We remove \emph{PSF0} and \emph{PSF1} event types from the analysis below 0.1\,GeV and keep all event types above 0.1\,GeV.  
\emph{PSF0} and \emph{PSF1} are events classified with poor PSF and removing these event types thereby improves the PSF with the trade-off of less data.  
This reduces the possibility of source confusion from nearby weak sources. 
This also reduces the chance of false positive detections as described in the \fermi{} low energy catalog (1FLE)\cite{2018A&A...618A..22P}.
% 20220216 JS removed
%The obtained SEDs are presented in Fig.~3.

\subsection{Optical photometry}

Optical photometric observations of \oph{} were carried out by Joan Or\'o Telescope (TJO) and {\it Asiago Novae \& Symbiotic stars Collaboration} (ANS, telescopes ID 310, 610 and 2203). The TJO is a 1-meter class robotic telescope located at Montsec observatory (42.05$^\circ$N, 0.73$^\circ$E), Catalonia, Spain. 
The multi band ({\it BVR$_c$I$_c$}) data were analysed using a semi-automatic pipeline for differential photometry \cite{2018A&A...620A.185N} assuming the aperture radius of 7.5$^{\prime\prime}$.
The comparison stars magnitudes are obtained from American Association of Variable Star Observers International Database (AAVSO). The stars are numbered as 115, 121, 129, 130, and 133 in the database finding chart.
% 20220216 JS removed
%The observed magnitude of the source is reported in Supplementary Table \ref{tab:magnitudes}.
%and the $V$-band light-curve is illustrated in Fig.~1 (bottom panel), as an example, together with the data obtained from American Association of Variable Star Observers International Database (AAVSO).

The data obtained by ANS are analyzed using PSF photometry method described in \cite{2012BaltA..21...13M, 2012BaltA..21...22M}. 
The same local photometric sequence, extracted from APASS DR8 all-sky survey \cite{2012JAVSO..40..430H, 2014CoSka..43..518H} and accurately placed on the system of equatorial standards \cite{2009AJ....137.4186L}  via the color equations calibrated in  \cite{2014JAD....20....4M, 2014AJ....148...81M}, has been used for all telescopes ensuing a high consistency of the data.  
The photometry results are given in Supplementary Table\,6%\ref{tab:magnitudes}
, where the quoted uncertainties are the total error, which quadratically combine the measurement error on the variable with the error associated to the transformation from the instantaneous local photometric system to the standard one (as defined by the photometric comparison sequence).  
All measurements were carried out with aperture photometry. 
% 20220216 JS removed
%The resulting light- and color-curves are plotted in Fig.\,\ref{figs:photometry}, where they are compared to the similar data for the 2006 event as published in \cite{2007BaltA..16...46M}.

The cross calibration between instruments was performed by using the color index of the source. 
The data obtained by two telescopes are in good agreement. 
However, to reduce the systematic uncertainties, minimal offsets ($B-V  = +0.03$, $V-R_c = +0.05$, and $V-I_c = -0.02$) were applied to TJO data. 
The contribution of the strongest emission lines ($H_{\alpha}$ and $H_{\beta}$) were removed from the observed magnitude using the simultaneous spectroscopic observations from the publicly available optical spectra in Astronomical Ring for Access to Spectroscopy (ARAS) \cite{2019CoSka..49..217T}. 
We found that the contribution of the $H_{\beta}$ emission line in the $V$-band is negligible for the first ten days after the outburst. 
The contribution of the $H_{\beta}$ emission line is significant in the $B$-band and increases from 3\% to 15\% during the same time interval.
Moreover, the contribution of the $H_{\alpha}$ emission line is dominant in the $R$-band and increases from 5\% to 83\% during the same time interval owing to a sudden jump from 5\% to 34\% between $T-T_0 = 0.98$\,days and $T-T_0 = 2.89$\,days. 
The results of these corrections are presented in Supplementary Table~7%\ref{tab:corr_mag}
. 

All optical data described in this section are corrected for the effect of Galactic extinction by assuming $E(B-V) = 0.65$ \cite{2014ApJ...785...97H}, Galactic extinction law \cite{1989ApJ...345..245C}, and the absolute fluxes (corresponding to zero magnitude) \cite{1998A&A...333..231B} in each band.

During the nova outburst the photosphere emission creates the dominant  radiation field.
We describe the radiation field using photometric and spectroscopic measurements by applying black body approximation. 
During the first four days of the nova, contemporaneous with the MAGIC measurements, the emission can be described by the photosphere temperature dropping from $T_{ph} = 10800$\,K to $7680$\,K and radius $R_{ph} = 200\,R_\odot$ (see EDF~6%\ref{figs:spect}
).
It should be noted that the asymmetry of the photosphere (see e.g. \cite{2007A&A...464..119C, 2015NewA...36..128S}, lack of measurements at the shortest wavelengths and the presence of lines affect the above mentioned fits. 
Therefore, the photosphere radius and temperature values should be considered only a crude approximation of the radiation field, in context of gamma-ray emission, and no conclusion on the evolution of those two parameters should be drawn. 
Noteworthly, the photosphere fit of 2006 eruption \cite{2015NewA...36..128S}, when rescaled to the nova distance of $2.45$\,kpc, provides a similar radius ($245 -310)\,R_\odot$, and temperature ($8200$\,K).

\subsection{Spectroscopy and ejecta kinematics}\label{sect:spectroscopy}
RS Oph spectra during the 2021 outburst have been acquired with the Echelle spectrograph of the Varese 0.84 m telescope \cite{2021arXiv210901101M} and the Catania Astrophysical Observatory Spectropolarimeter \cite{2016AJ....151..116L} of the Catania 0.91 m telescope. 
%and HARPS-North \cite{2012SPIE.8446E..1VC} of the Telescopio Nazionale Galileo. 
The reduction of spectra, which included the subtraction of the bias frame, trimming, correcting for the flat-field and the scattered light, extraction for the orders, and wavelength calibration, was done as in \cite{2015MNRAS.451..184C} by using the NOAO/IRAF packages.
IRAF is distributed by the National Optical Astronomy Observatory, which is operated by the Association of Universities for Research in Astronomy, Inc.
% \footnote{IRAF is distributed by the National Optical Astronomy Observatory, which is operated by the Association of Universities for Research in Astronomy, Inc.}
% 20220216 JS removed
%The log-book of observations is in Table\,\ref{logbook_spectroscopy}. 

%\subsubsection{Kinematics}
The H$_{\alpha}$ profile obtained on day $T-T_0 = 0.91$ consists of a triangular shape with Full Width at Zero Intensity of $\sim7500$\,km\,s$^{-1}$ and a blue shift absorption component at $4250$\,km\,s$^{-1}$, exactly as it was reported by \cite{2008ASPC..401..227S} $1.38$ days after the 2006 outburst of RS Oph. %FWZI

The close similarity of the 2006 and 2021 spectral line profiles along the envelope expansion is testified on day $T-T+0 = 15$ by the presence of satellite components at the same high-velocity ($2500$\,km\,s$^{-1}$). 
This feature was associated by \cite{2008ASPC..401..227S} %Skopal and coworkers 
to a presence of two jets (c.f. figs.~1 and 2 therein and EDF~7%\ref{figs:line_profiles}
). 
Also, \cite{2018MNRAS.474.4211M} measured a velocity of $4200$\,km\,s$^{-1}$ the day after the outburst.

Because of the day-by-day changing of absorption and emission features across the whole \oph{} spectrum, we have determined the velocity of the expanding envelope as the terminal value simultaneously representative
of the H$_\alpha$, H$_\beta$ and He\,I\,5876$\lambda$ P-Cygni profiles (Supplementary Table\,8%\ref{logbook_spectroscopy}
). 
An error of 250 km\,s$^{-1}$ was assumed as representative of differences between profiles. 
% 20220216 JS removed
%Results are in Table\,\ref{logbook_spectroscopy}.
EDF~7 %\ref{figs:line_profiles}
shows these profiles in the first three days after the expansion as well as on days 5 and 15.  
% 20220216 JS removed
%The expansion velocity in time is given in the bottom right panel of Fig.~\ref{figs:line_profiles}.

The acceleration along the initial three days is not statistically confirmed and we assume (4500 $\pm$ 250)\,km\,s$^{-1}$ as representative of the ejecta expansion at the earliest stage (during the VHE gamma-ray detection by MAGIC).

It is worth to remind that this velocity is volume average, weighted by the brightness, temperature and density of the ejecta velocities and agrees with results from the modeling by \cite{2009ApJ...703.1955R} of the HST images of the spatially resolved and expanding ejecta during the 2006 event.
Radio maps of the 2006 outburst of RS Oph \cite{2008ApJ...685L.137S} have shown the presence of highly collimated flows with a velocity close to 10000 km\,s$^{-1}$.  %at the Gaia eDR3 distance of 2.67 kpc,
In this framework, the decrement of the velocity after the initial days 
%with a linear trend, $V = 2990-27.6 (T-T_0)\mathrm{[km\,s^{-1}]}$,
is simply a consequence of a non-spherical mass outflow \cite{2008ASPC..401..227S, 2007ApJ...665L..63B, 2009ApJ...703.1955R}.

%\clearpage
\section{Modeling}\label{sec:model}

There are compelling both simulation (see e.g. \cite{2016MNRAS.457..822B}) and observational (see e.g. \cite{2007ApJ...665L..63B}) evidence that the mass transfer in symbiotic binaries causes non-spherical circumstellar environment.
Such asymmetries are crucial when considering the morphology of the emission in particular in optical and X-ray ranges. 
Here, using a similar approach to \cite{2012PhRvD..86f3011S,2015A&A...582A..67A},  we consider a simplified, spherically-symmetric scenario in order to evaluate the conditions in which gamma-ray radiation can be produced by either electrons or protons and to investigate spectral features of such an emission. 
The used parameters are summarized in Supplementary Table~10%\ref{tab:nova_par}%. 

\subsection{Acceleration and cooling of particles}\label{sect:maxe}
We parametrize the acceleration of charged particles with acceleration parameter $\xi$:
\begin{equation}
\left(\frac{dE}{dt}\right)_{acc} = \frac{\xi c E}{R_L(E)}, %% = 0.90 (\xi/10^{-4}) (B/\mathrm{G}),
\end{equation}
where $R_L(E)$ is the Larmor radius of particle with energy $E$ in perpendicular magnetic field $B$. 
The corresponding acceleration time scale, expressed in days, can be computed as:
\begin{equation}
t_{acc} = E/\left(\frac{dE}{dt}\right)_{acc} = 3.9 \left(\frac{E}{\mathrm{300 \,GeV}}\right) \left(\frac{\xi B}{10^{-7}\,\mathrm{G}}\right)^{-1}\, \mathrm{[day]}.
\end{equation}
The maximum achieved energies will stem from balancing such acceleration time with ballistic time $t_{bal}$, defined as the time from the onset of the nova, or by dominating cooling process. 
The shock distance $R_{sh}$ at the time $t = T-T_0$ can be estimated based on its speed $v_{sh}$:
\begin{equation}
    R_{sh} = 1.2\times 10^{14} \left(\frac{v_{sh}}{4500\,\mathrm{km\,s^{-1}}}\right) \left(\frac{t}{3\,\mathrm{d}}\right)\,\mathrm{[cm]}. %% 1.17
\end{equation}
As the nova shock expands the adiabatic energy losses will be directly connected with the ballistic time. 
We define the adiabatic time scale as the time in which the energy of particles decreases by a factor of $e$, resulting in $t_{adiab} = e\,t_{bal}$.

The protons will cool on hadronic interactions with the ambient matter, either the nova ejecta, or the RG wind. 
We assume that the ejecta concentrate at the distance of $R_{sh}$ in a layer with a thickness of $h\times R_{sh}$, with $h = 0.1$.
The number density of the ejecta can be estimated as:
\begin{equation}
    n_{ej} = \frac{M_{ej}}{4 \pi h R_{sh}^3 m_p} = 
    6.0\times10^8 \frac{M_{ej}}{10^{-6} M_\odot} %5.96
    \left (\frac{v_{sh}}{4500\,\mathrm{km\,s^{-1}}}\right)^{-3} \left(\frac{t}{3\,\mathrm{d}}\right)^{-3}
    \left(\frac{h}{0.1}\right)^{-1}\, \mathrm{[cm^{-3}]},
\end{equation}
where $M_{ej}$ is the total ejected mass and $m_p$ is the proton mass.
Alternative assumption that the ejecta fill homogenously a sphere with radius $R_{sh}$ would result in a factor of 3 lower value of $n_{ej}$. 
The number density of the ambient material in the RG wind can be estimated as: 
\begin{eqnarray}
    n_{RG} &=& \frac{\dot M_{RG}}{4 \pi R_{sh}^2 v_{RG} m_p} \\\nonumber
    &=& 1.1\!\times\!10^8 \frac{\dot M_{RG}}{5\times10^{-7} M_\odot / \mathrm{yr}}\!
    \left(\frac{v_{sh}}{4500\,\mathrm{km\,s^{-1}}}\right)^{\!-2}\!\! \left(\frac{t}{3\,\mathrm{d}}\right)^{\!-2}\!\!
    \left(\frac{v_{RG}}{10\,\mathrm{km\,s^{-1}}}\right)^{\!-1}\! \mathrm{[cm^{-3}]},
\end{eqnarray}    
where $v_{RG}$ is the speed of the RG wind and $\dot M_{RG}$ is the mass loss rate of the RG. 
The total density of the ambient medium for the hadronic interaction for the assumed parameters of \oph{} it is mostly dominated by the ejecta ($n_p \approx n_{ej}$). 
The proton cooling time scale on hadronic p-p interactions can be then computed as:
\begin{equation}
    t_{pp} = (n_p c \sigma_{pp})^{-1} = 21 (n_p/6\times10^8\,\mathrm{cm^{-3}})^{-1}\,\mathrm{[day]} ,
\end{equation}
where $\sigma_{pp} = 3\times 10^{-26}$\,cm$^2$.
As the cooling timescale is longer than the ballistic time, the maximum energies to which protons can be accelerated are determined by the time from the nova onset. 

In the case of electrons, cooling losses can originate either from inverse Compton scattering on the photosphere thermal radiation or from Bremsstrahlung radiation on the ambient matter.
We compute the inverse Compton cooling time scale taking into account Klein-Nishina correction factor following \cite{2005MNRAS.363..954M}
\begin{equation}
    t_{IC} = \frac{3(m_ec^2)^2}{4c\sigma_T u_{ph} E}\left(1+4\epsilon_{ph} E/(m_ec^2)^2\right)^{1.5},
\end{equation}
where $m_e$ is the electron mass. The total energy density, $u_{ph}$, and characteristic temperature of soft photons, $\epsilon_{ph}$, can be estimated as 
\begin{eqnarray}
u_{ph}& = &0.14\frac{(R_{ph}/200\,R_\odot)^2 (T_{ph}/8460\,\mathrm{K})^4}
{(v_{sh}/4500\,\mathrm{km\,s^{-1}})^2 (t/3\,\mathrm{d})^2} \mathrm{[erg\, cm^{-3}]} \\ % 0.138
\epsilon_{ph}& = &2.2(T_{ph}/8460\,\mathrm{K})\,\mathrm{[eV]}. % 2.18
\end{eqnarray}
For the used above scaling values the dependence of $t_{IC}$ with energy can be described as 
$t_{IC} = 4.4 \times 10^{-3} (E/300\,\mathrm{GeV})^{-1})[1+10(E/300\,\mathrm{GeV})]^{1.5}\,\mathrm{[day]}$
resulting in fast cooling of high-energy electrons. 
We estimate the Bremsstrahlung losses using the same density of ambient matter $n_p$ as 
\begin{equation}
    t_{brems} = X_0/(n_p m_p c ) = 24 (n_p/6\times10^8\,\mathrm{cm^{-3}})^{-1}\,\mathrm{[day]} ,
\end{equation}
where $X_0 = 63\,\mathrm{g\,cm^{-2}}$ is the radiation length in proton gas.
For the expected parameters of \oph{}, the Bremsstrahlung losses are thus negligible. 
Also the synchrotron energy losses are negligible, unless the magnetic field in the shock reaches the level of about $1$\,G. 

% 20220216 JS removed
%The different time scales are summarized in Fig.~\ref{figs:timescales}.
%
In order to accelerate protons up to energies of a few hundred GeV, the value of $\xi B \gtrsim 10^{-7}$\,G is required (EDF~5%\ref{figs:timescales}
). 
If electrons are accelerated in the same conditions, they can reach energies of only $\sim10$\,GeV.
In order to explain the observed gamma-ray emission reaching hundreds of GeV, much higher values $\xi B \gtrsim 3\times 10^{-6}$\,G are required.
Second-order Fermi acceleration on the nova shock is expected to provide acceleration parameter of the order of $\xi \lesssim (v_{sh}/c)^2 \approx 10^{-4}$, resulting in the requirement of $B\gtrsim0.03$\,G fields for the electron case and much weaker $B\gtrsim$\,mG for the proton one. 

\subsection{Energetics}\label{sect:energy}

The kinetic energy of the ejecta can be estimated as:
\begin{equation}
    E_k = 0.5 M_{ej}v_{sh}^2 = 2.0\times 10^{44} \left(\frac{M_{ej}}{10^{-6}M_\odot}\right)\left(\frac{v_{sh}}{4500\,\mathrm{km\,s^{-1}}}\right)^2\,\mathrm{erg}
\end{equation}
For the assumed parameters determining the density of target material, the fit of the proton energy distribution in Fig.~3 requires a total power in protons of $4.4 \times 10^{43}$\,erg. 
This energetics requirement scales with the assumed model parameters as:
\begin{equation}
    E_{p,nova} = 0.44 \times 10^{44} \left(\frac{M_{ej}}{10^{-6}M_\odot}\right)^{-1}\left(\frac{v_{sh}}{4500\,\mathrm{km\,s^{-1}}}\right)^{3}\left(\frac{d}{2.45\mathrm{kpc}}\right)^{-2}\frac{h}{0.1}\,\mathrm{erg}
\end{equation}
Therefore the efficiency of conversion of energy from the shock to protons can be computed as: % the ratio of the above two formulae:
\begin{equation}
    \epsilon = \frac{E_{p,nova}}{E_k} = 0.22  \left(\frac{M_{ej}}{10^{-6}M_\odot}\right)^{-2}\left(\frac{v_{sh}}{4500\,\mathrm{km\,s^{-1}}}\right) \left(\frac{d}{2.45\mathrm{kpc}}\right)^{-2}\frac{h}{0.1}
\end{equation}
It is clear that protons need to obtain a significant fraction ($\sim20\%$) of the shock kinetic energy.
Lower fraction could be achieved if the mass of the ejecta is higher, it is more concentrated at the shock (lower $h$) or if the speed of the shock is decreased. 
Concentration of the nova ejecta and proton acceleration in the bipolar direction would increase the target material density and efficiency of the gamma ray production.
This would further lower the total energy required in the accelerated protons compared to the assumed here spherically symmetric scenario.

\subsubsection{Contribution to the Cosmic Ray sea}\label{sect:cr_sea}

These accelerated protons eventually escape the nova to be part of the sea of Cosmic Rays. Since they do not suffer strong energy losses due to their interaction with intergalactic magnetic and photon fields, as it is the case for electrons, their contribution may extend to large distance from the nova explosion at all energies. 
Assuming that the energy released in all novae into accelerated protons is similar to that released in \oph{} ($E_{p,nova} = 4.4 \times 10^{43}$ erg) and a nova rate of $\sim50$ per year \cite{2017ApJ...834..196S} we get a total of  %$50^{+31}_{-23}$
\begin{equation}
    {\rm Novae\ energy\ rate} = E_{p,nova} \times {\rm nova\ rate} = 2.2 \times 10^{45} \mathrm{[erg/year]}
\end{equation}
It is considered that a supernova explosion usually releases $E_{\rm SN}\sim10^{51}$ erg \cite{2015MNRAS.451.2757W}, out of which $\sim$10\% can be converted into accelerated protons at the shock between the supernova ejecta and the interstellar medium (ISM). The SN rate in the galaxy is $\sim$2 per century \cite{2021NewA...8301498R}, therefore the supernova energy rate would be:
\begin{equation}
    {\rm Supernovae\ energy\ rate} = 0.1 \times E_{\rm SN} \times {\rm supernova\ rate} = 2 \times 10^{48} \mathrm{[erg/year]}
\end{equation}
making the contribution of novae $\lesssim$0.2\% to that of supernovae.

Let us now assume that the average energy density in CRs in the Milky Way is $E_{\rm dens, CRs}\sim$1.8 eV/cm$^3$ \citep{Webber_1998}. We would like to compute what is the region in which the energy density of the protons accelerated by the nova dominates over this energy density. The energy density of these protons will be given by the total energy ($E_{p,nova}$) divided by the volume of the region
\begin{equation}
    {E_{\rm dens, nova, 1\,eruption}} = \frac{3 E_{p,nova}}{4 \pi R_{\rm eruption}^3} 
\end{equation}
where $R_{\rm eruption}$ is the radius of the region. If we compare ${E_{\rm dens, nova}} = E_{\rm dens, CRs}$, we obtain $R_{\rm eruption}\sim$0.5 pc, that is subject to the assumption on the energy density performed and may change if larger energy densities are considered \citep{2020MNRAS.497.1712B}.

Finally, in the special case of a recurrent nova like \oph{} that repeats its explosions every $\sim$15 years \citep{2009ApJ...697..721S}, we would get this energy injection repeated over time. Considering a period of recurrence of up to $10^5$ years, the region over which this nova would dominate has a size of:
\begin{equation}
    {E_{\rm dens, nova, recurrent}} = \frac{3 E_{p,nova} \times 10^4}{4 \pi R_{\rm recurrent}^3} 
\end{equation}
and the radius over which the protons accelerated by the nova would dominate over the energy density of the ISM would be $R_{\rm recurrent}\sim$9 pc.

%\clearpage

\hspace{1em}

\noindent {\bf Availability of data and materials:} Analysis products of MAGIC data are available here: \url{http://vobs.magic.pic.es/fits/}. Low level data are available on request. 

\noindent {\bf Code availability:} The code for fitting the electron and proton models is available in \url{https://opendata.magic.pic.es/download?pid=2}.
%\section*{Declarations}

\bmhead{Acknowledgments}

We would like to thank the Instituto de Astrof\'{\i}sica de Canarias for the excellent working conditions at the Observatorio del Roque de los Muchachos in La Palma. The financial support of the German BMBF, MPG and HGF; the Italian INFN and INAF; the Swiss National Fund SNF; the ERDF under the Spanish Ministerio de Ciencia e Innovaci\'{o}n (MICINN) (PID2019-104114RB-C31, PID2019-104114RB-C32, PID2019-104114RB-C33, PID2019-105510GB-C31,PID2019-107847RB-C41, PID2019-107847RB-C42, PID2019-107847RB-C44, PID2019-107988GB-C22); the Indian Department of Atomic Energy; the Japanese ICRR, the University of Tokyo, JSPS, and MEXT; the Bulgarian Ministry of Education and Science, National RI Roadmap Project DO1-400/18.12.2020 and the Academy of Finland grant nr. 320045 is gratefully acknowledged. This work was also supported by the Spanish Centro de Excelencia ``Severo Ochoa'' (SEV-2016-0588, SEV-2017-0709, CEX2019-000920-S), the Unidad de Excelencia ``Mar\'{\i}a de Maeztu'' (CEX2019-000918-M, MDM-2015-0509-18-2) and by the CERCA program of the Generalitat de Catalunya; 
by the Croatian Science Foundation (HrZZ) Project IP-2016-06-9782 and the University of Rijeka Project uniri-prirod-18-48; % updated
%by the Croatian Science Foundation (HrZZ) Project IP-2016-06-9782 and the University of Rijeka Project 13.12.1.3.02; %old
by the DFG Collaborative Research Centers SFB823/C4 and SFB876/C3; the Polish National Research Centre grant UMO-2016/22/M/ST9/00382; and by the Brazilian MCTIC, CNPq and FAPERJ.
%TJO
The Joan Oró Telescope (TJO) of the Montsec Observatory (OdM) is owned by the Catalan Government and operated by the Institute for Space Studies of Catalonia (IEEC). 
%AAVSO
We acknowledge with thanks the variable star observations from the AAVSO International Database contributed by observers worldwide and used in this research.
% CAOS
We gratefully acknowledge the prompt response to the alert and the data provided by the CAOS Team. 
%ARAS
We acknowledge with thanks the Astronomical Ring for Amateur Spectroscopy (ARAS) database \cite{2019CoSka..49..217T} (\url{https://aras-database.github.io/database/index.html}). The observers who contributed worldwide and used in this research are Olivier Garde,	 Vincent Lecoq, Lorenzo Franco, Francois Teyssier, Olivier Thizy, Christophe Boussin, Pavol A. Dubovsky, and David Boyd.
The authors would like to thank Giacomo Principe for the advice in extending the \fermi{} analysis below 100 MeV and Filippo D'Ammando for his comments on the manuscript.
R.L-C.'s work was financially supported by the European Union's Horizon 2020 research and innovation program under the Marie Skłodowska-Curie grant agreement No. 754496 - FELLINI.
We would like to thank the anonymous journal reviewers for the comments that helped to improve the manuscript. 

\noindent {\bf Author Contribution Statement:} 
The individual authors who contributed to this manuscript in alphabetic order are 
W. Bednarek: theoretical interpretation; 
V. Fallah Ramazani: analysis and coordination of the optical photometry data, drafting of the corresponding paper section; 
D. Green: trigger of the MAGIC observations, analysis of the MAGIC data, drafting and edition of the manuscript; 
F. Leone: coordination and analysis of the optical spectroscopy data, interpretation of ejecta kinematics; 
R. L\'opez-Coto: analysis of the MAGIC and Fermi-LAT data, theoretical interpretation, comparison with other novae, computation of the contribution to CRs, drafting and edition of the manuscript; 
A. L\'opez-Oramas: trigger and coordination of the MAGIC campaign, analysis of the MAGIC data, drafting and edition of the manuscript; 
U. Munari: analysis of the optical photometry data and cross-calibration of the different optical instruments; 
J. Sitarek: coordination of the MAGIC novae observation program, analysis of the MAGIC data, theoretical modelling, leadership of the publication effort, drafting and edition of the manuscript; 
P. Valisa: collection and analysis of the optical photometry data. 
The rest of the authors have contributed in one or several of the following ways: design, construction, maintenance and operation of the instrument(s) used to acquire the data; preparation and/or evaluation of the observation proposals; data acquisition, processing, calibration and/or reduction; production of analysis tools and/or related Monte Carlo simulations; discussion and approval of the contents of the draft.

\noindent {\bf Competing Interests:} The authors declare that they have no competing interests.

\section*{Figure Legends/Captions}
% THE ACTUAL FIGURES ARE COMMENTED OUT AS ONLY THE CAPTIONS SHOULD BE SUBMITTED AND FIGURES ARE SUBMITED AS SEPARATE FILES
%
% FIG1
\begin{figure}[h!]
    \centering
    \includegraphics[width=0.7\textwidth]{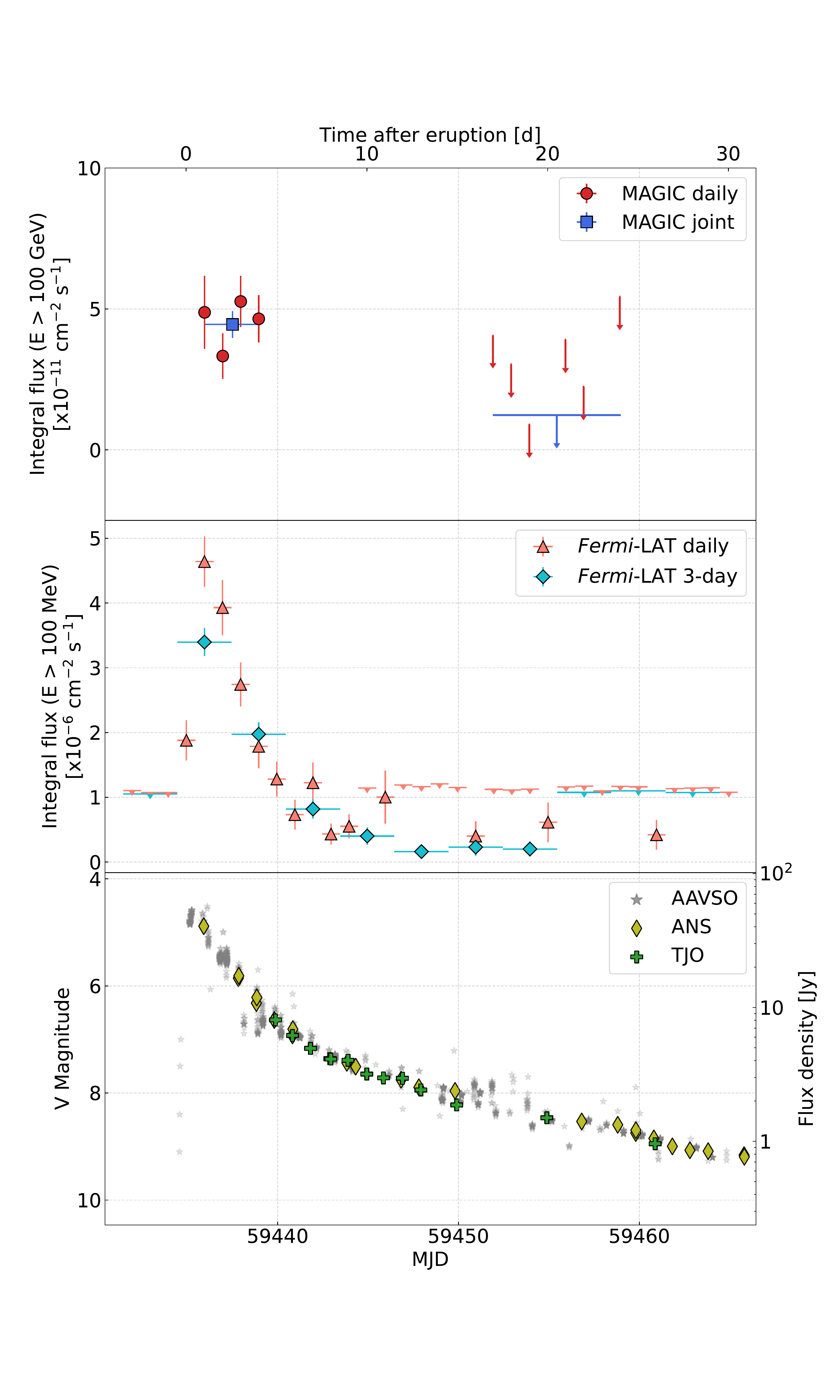}

    \caption{Multiwavelength light curve in the VHE (MAGIC, top panel), high-energy (\fermi{}, middle panel) and optical (TJO, ANS, and AAVSO, bottom panel) bands. The lack of MAGIC data between MJD 59440 and MJD 59454 is due to the presence of bad weather conditions and strong moonlight. Errorbars represent 1-sigma statistical uncertainties in the data points.}

    \label{fig1}
\end{figure}

% FIG2
\begin{figure}[h!]
    \centering
    \includegraphics[width=0.8\textwidth]{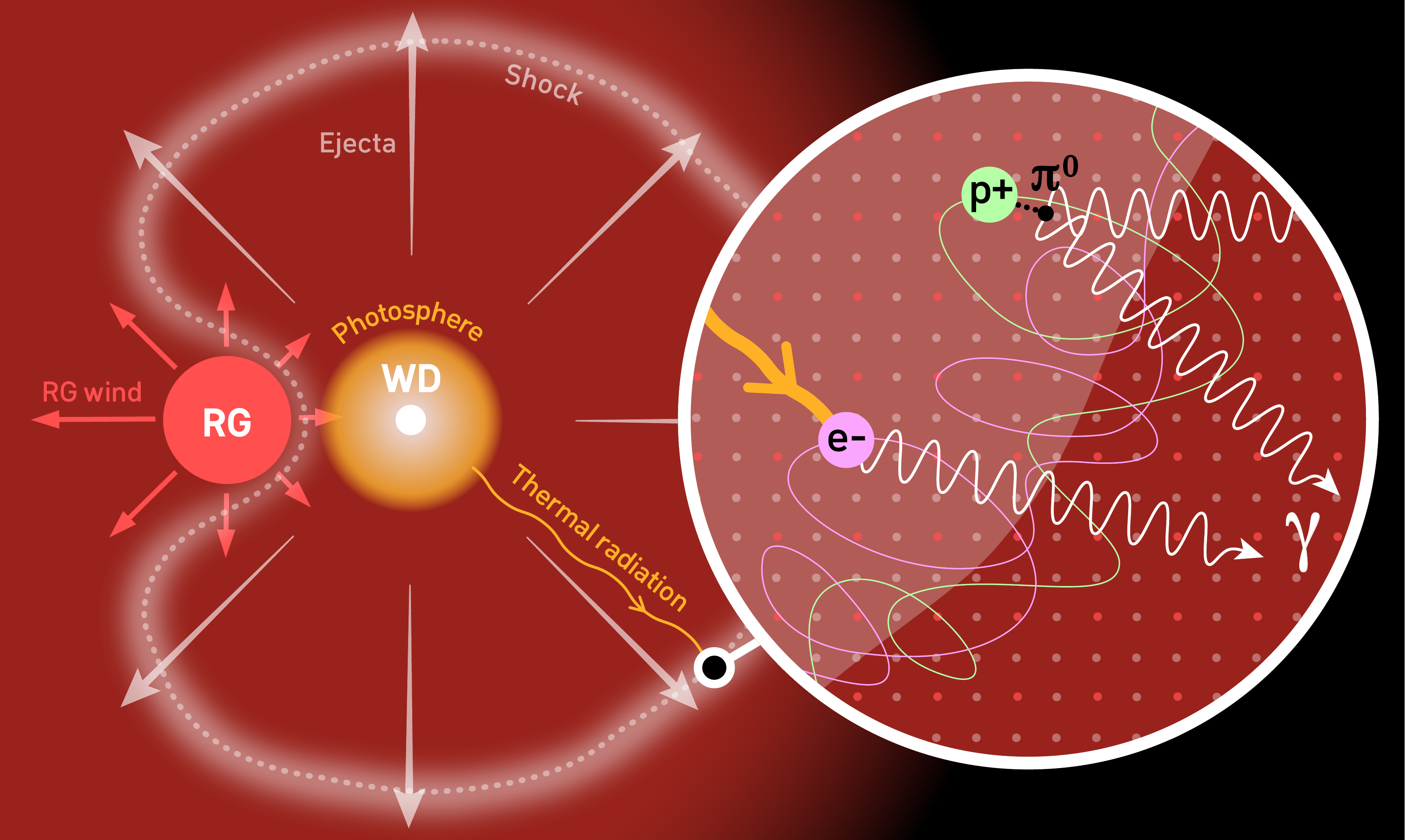}
    \caption{Schematic representation of \oph{} during an outburst. 
    A photosphere (yellow circle) surrounds the White Dwarf (WD, white small circle). 
    Its companion star, a red giant (RG, red circle) emits a slow wind (red arrows). 
    Ejecta of the nova explosion (gray arrows) propagate into the surrounding medium causing a shock wave encompassing the binary system (gray dashed line).
    In the shock wave, energetic electrons and protons  (magenta and green wavy lines, respectively) are trapped by a magnetic field and accelerated.
    Gamma rays (white arrows) are produced by either electrons scattering the thermal radiation of the photosphere (yellow arrow) or by protons interacting with the surrounding matter (gray and red dots).
    }
    \label{fig2}
\end{figure}
\begin{figure}[h!]
    \centering
    \includegraphics[width=0.49\textwidth]{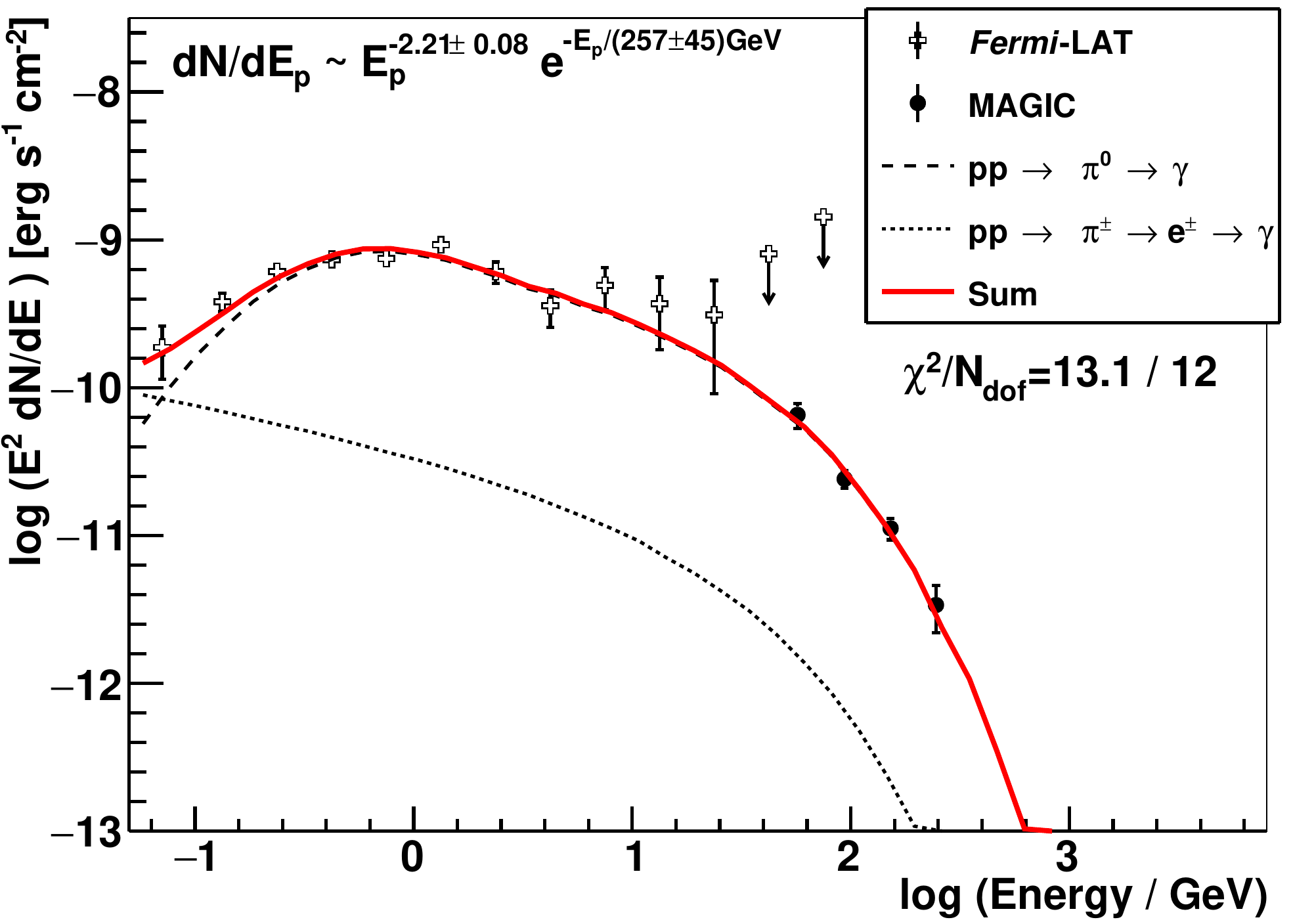}
    \includegraphics[width=0.49\textwidth]{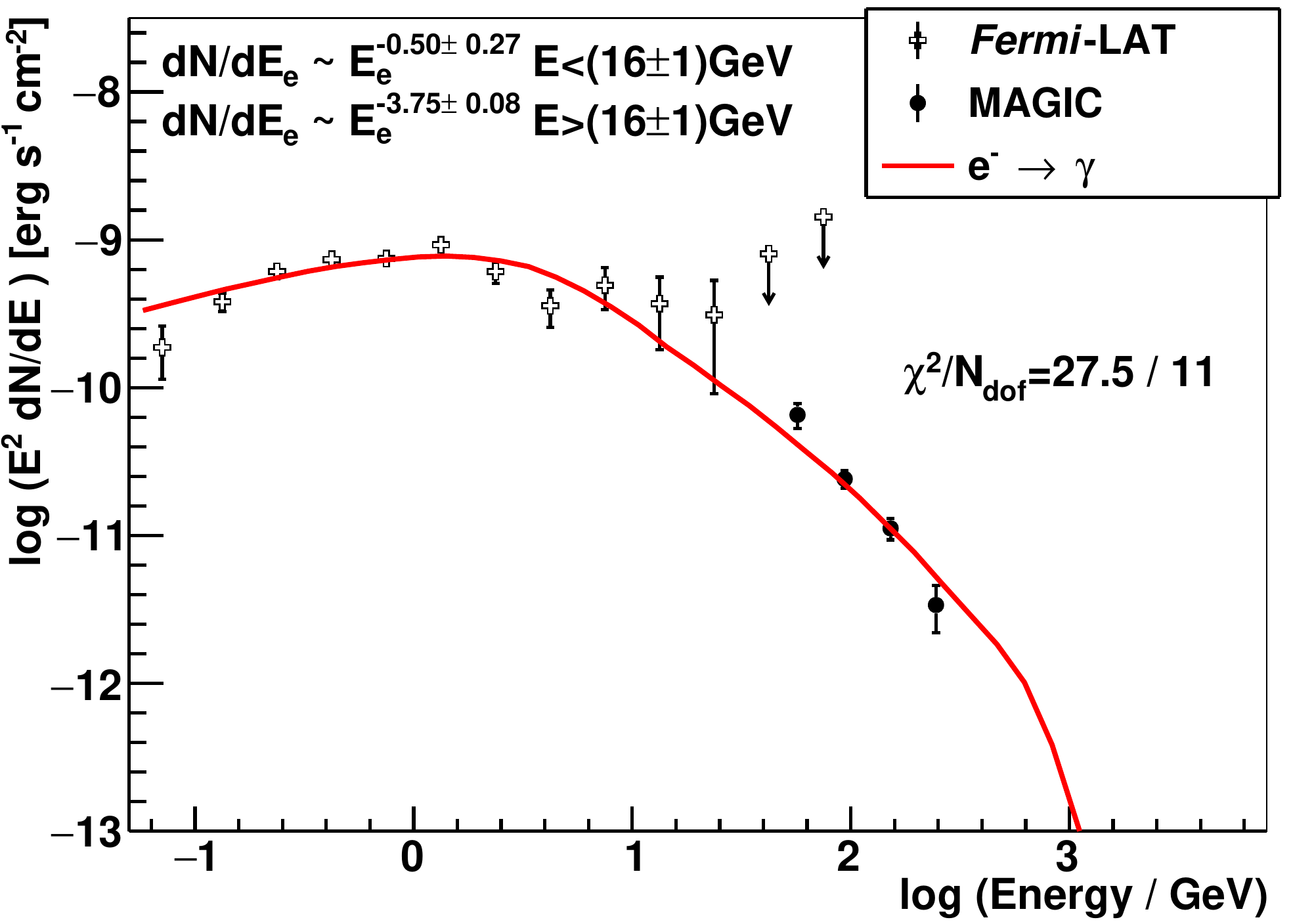}
    \includegraphics[width=0.49\textwidth]{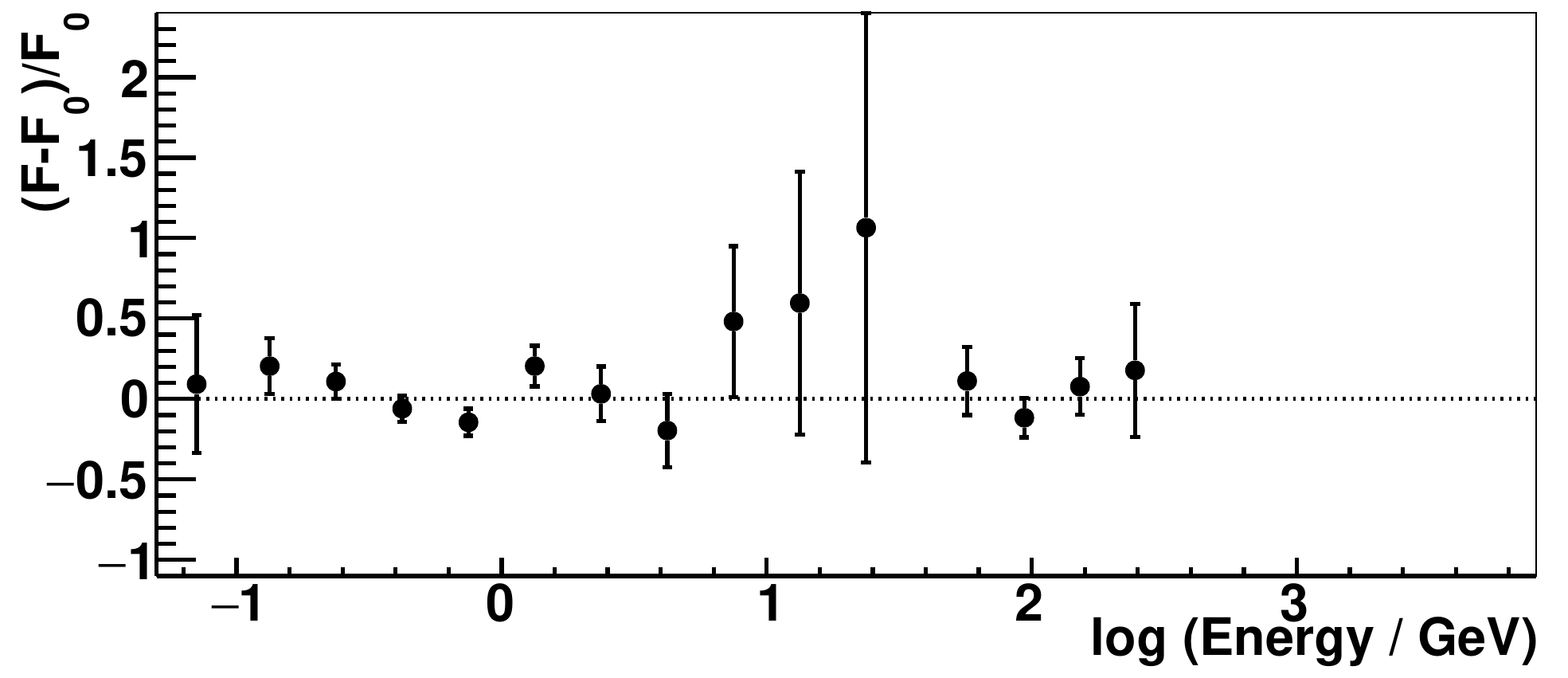}
    \includegraphics[width=0.49\textwidth]{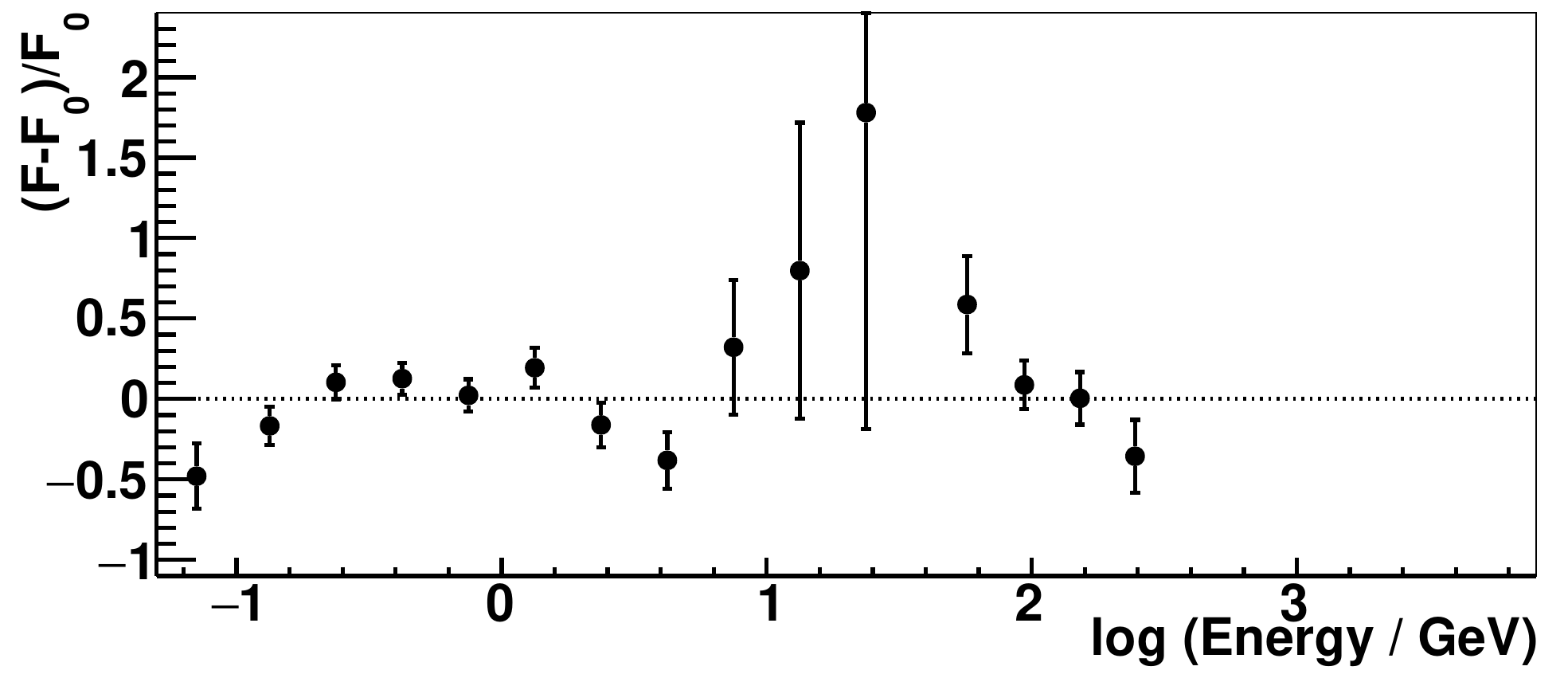}
    \caption{Gamma-ray spectrum of \oph\ observed with \fermi\ (empty crosses) and MAGIC (filled circles) averaged over the first four days of the outburst, modeled within hadronic (left panel) or leptonic (right panel) scenario.
    The dashed line shows the gamma rays from the $\pi^0$ decay and the dotted line shows the inverse Compton contribution of the secondary $e^\pm$ pairs produced in hadronic interactions.
    $dN/dE_p$ and $dN/dE_e$ report the shape of the proton and electron energy distributions obtained from the fit. 
    The bottom panel shows the fit residuals. Errorbars represent 1-sigma statistical uncertainties in the data points.
    }
    \label{fig3}
\end{figure}

% FIG4
\begin{figure}[h!]
    \centering
    \includegraphics[width=0.8\textwidth]{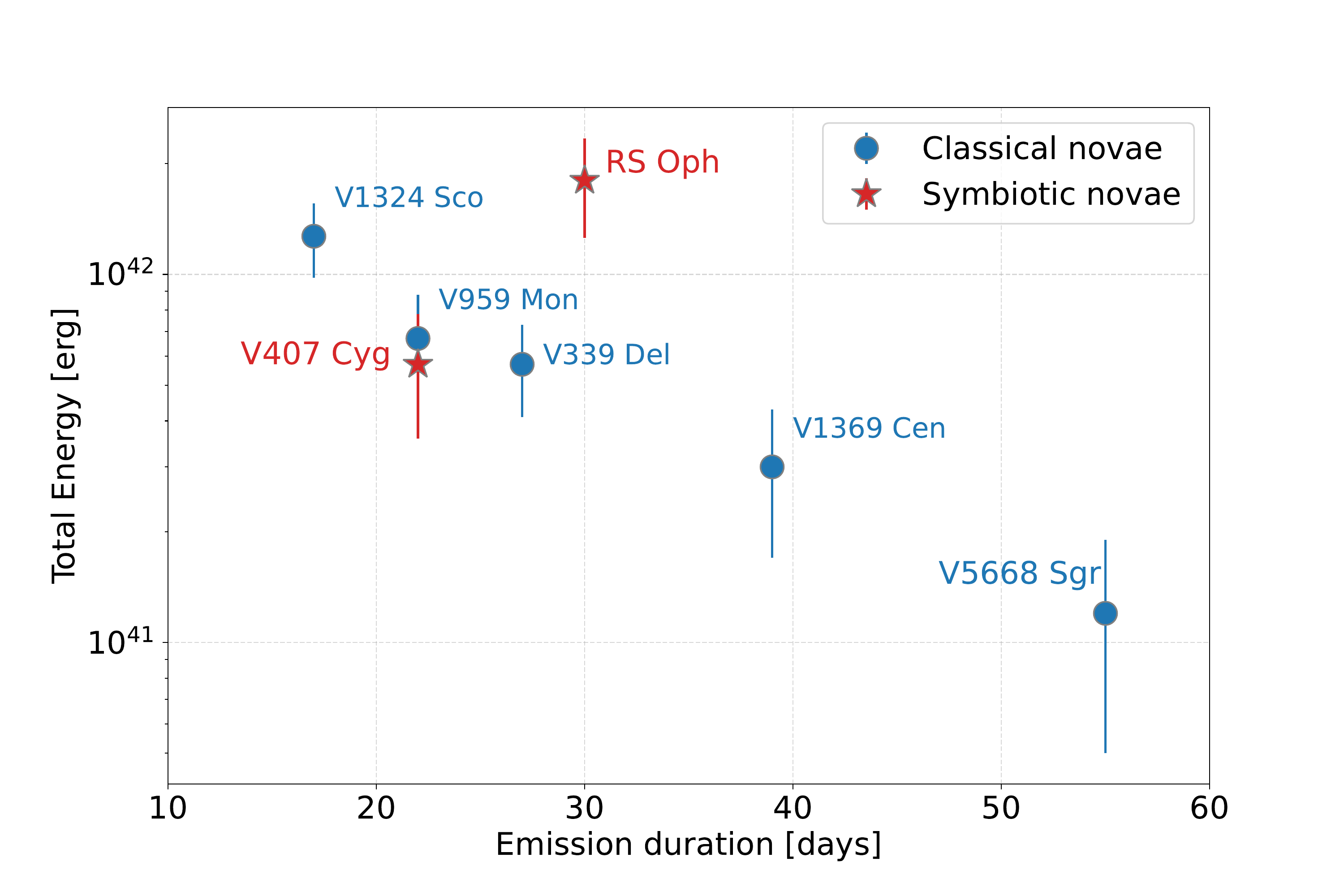}
    \caption{Total energy vs duration of \oph{} 2021 outburst compared to that of the other novae detected by \fermi{}. Data taken from \cite{2010Sci...329..817A, 2014Sci...345..554A, 2016ApJ...826..142C}. Errorbars represent 1-sigma statistical uncertainties in the data points.}
    \label{figs:novae_energy_photons}
\end{figure}
\clearpage

†\clearpage
\renewcommand{\figurename}{Extended Data Figure EDF} 
\setcounter{figure}{0}

% Extended data Figure 1
\begin{figure}[p]
    \centering
\includegraphics[width=\textwidth]{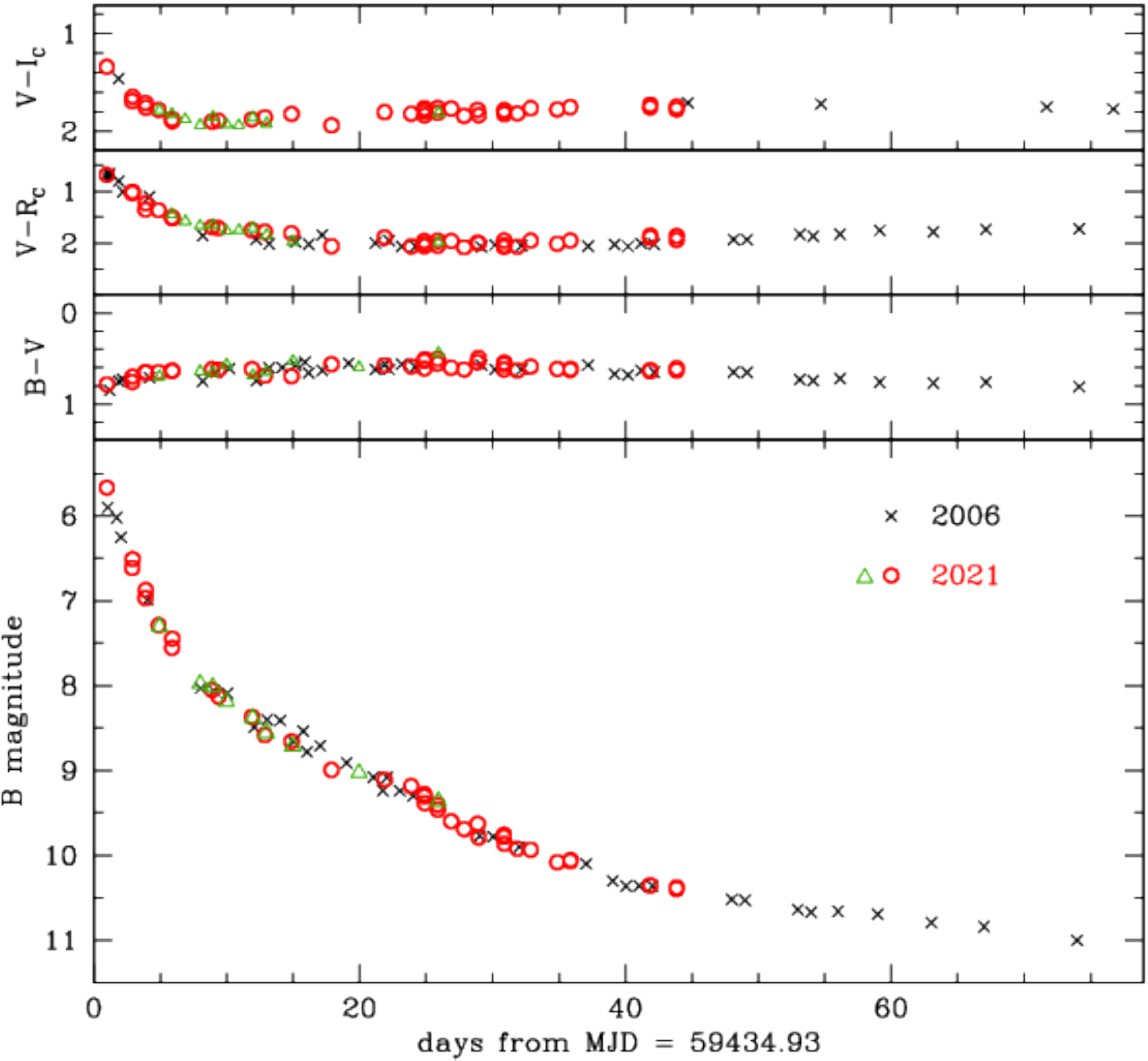}
\caption{Optical $B$-band observed magnitude and the color index of RS Oph 2021 outburst from ANS (red circles) and TJO (green triangle) compared to that of 2006 eruption (black crosses, computed with respect to MJD of $53775.86$, \cite{2007BaltA..16...46M}).
Top three panels show the color indices, while the bottom panel shows B magnitude evolution.}
\label{figs:photometry}
\end{figure}

% Extended data Figure 2
%
\begin{figure}[p]
    \centering
    \includegraphics[width=0.6\textwidth]{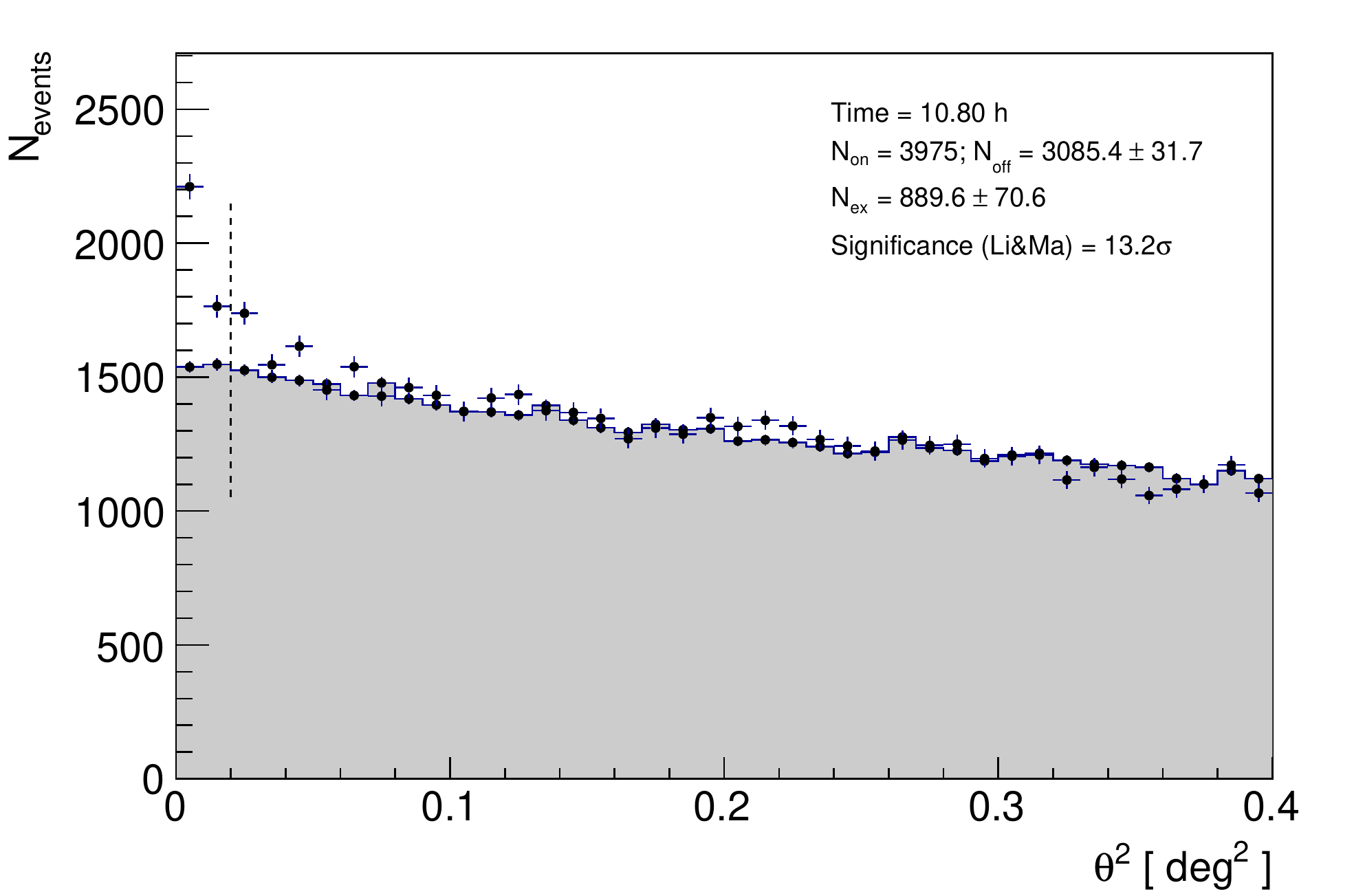}
    \caption{Distribution of the squared angular distance between the nominal source position and the reconstructed arrival direction of events (black crosses) and the estimated background (gray shaded area). 
    Vertical dashed line represents the angular cut below which the number of background and excess events as well as the statistical significance of the detection are given (inset panel). Error bars represent 1-sigma statistical uncertainties in the data points.}
    \label{figs:theta2}
\end{figure}
%

% Extended data Figure 3

%
\begin{figure}[t]
    \centering
    \includegraphics[trim=30 0 0 0, clip]{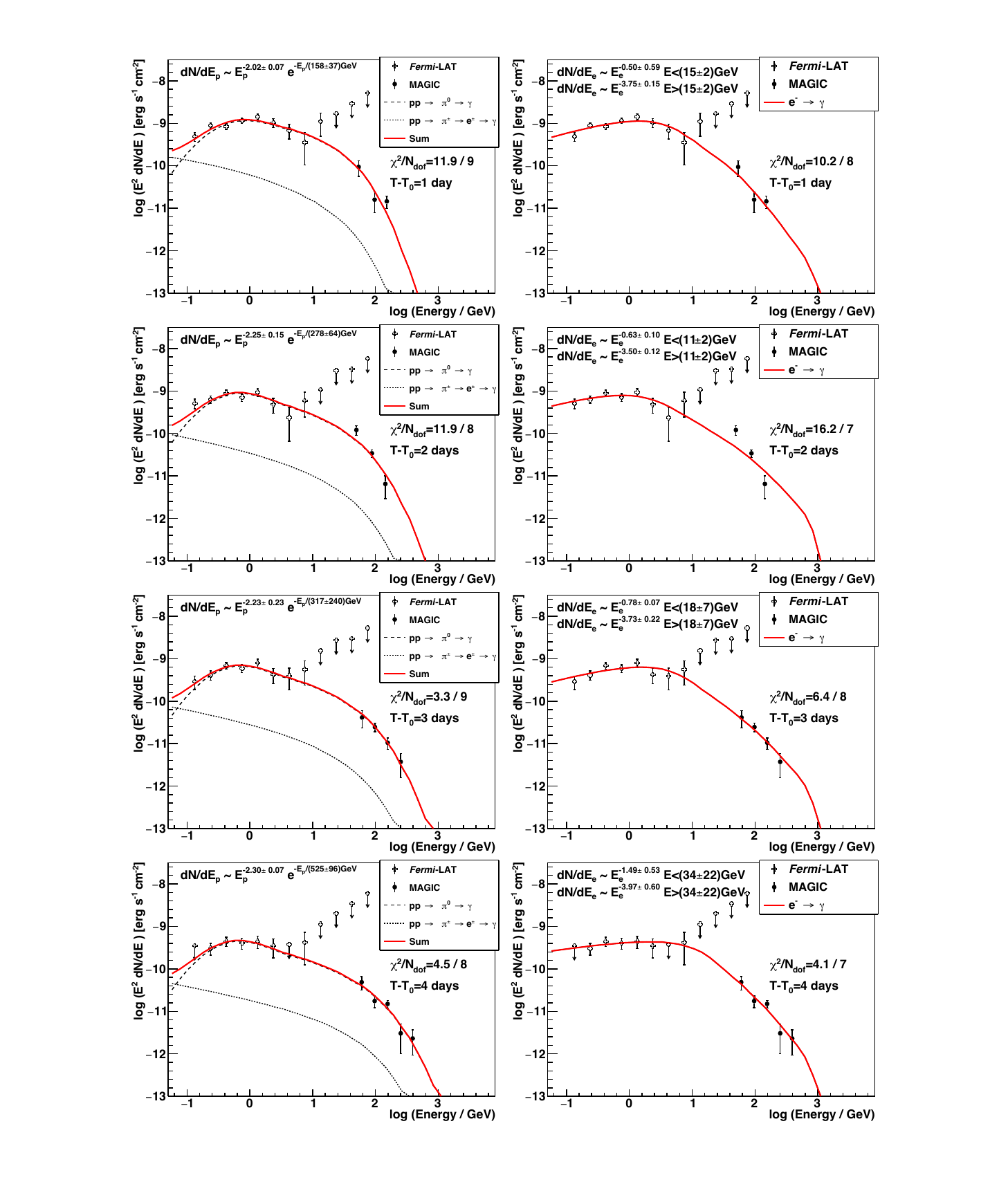}
    \caption{Modeling of daily emission in proton model (left panels) and electron model (right panels) for first, second, third and fourth day after the nova eruption (from top to bottom). 
    The dashed line shows the gamma rays from the $\pi^0$ decay and the dotted line shows the inverse Compton contribution of the secondary $e^\pm$ pairs produced in hadronic interactions.
    $dN/dE_p$ and $dN/dE_e$ report the shape of the proton and electron energy distributions obtained from the fit. 
    The bottom panel shows the fit residuals.
    Error bars represent 1-sigma statistical uncertainties in the data points.}
    \label{figs:sed_daily}
\end{figure}

% Extended data Figure 4
\begin{figure}[t]
    \centering
    \includegraphics[width=0.99\textwidth]{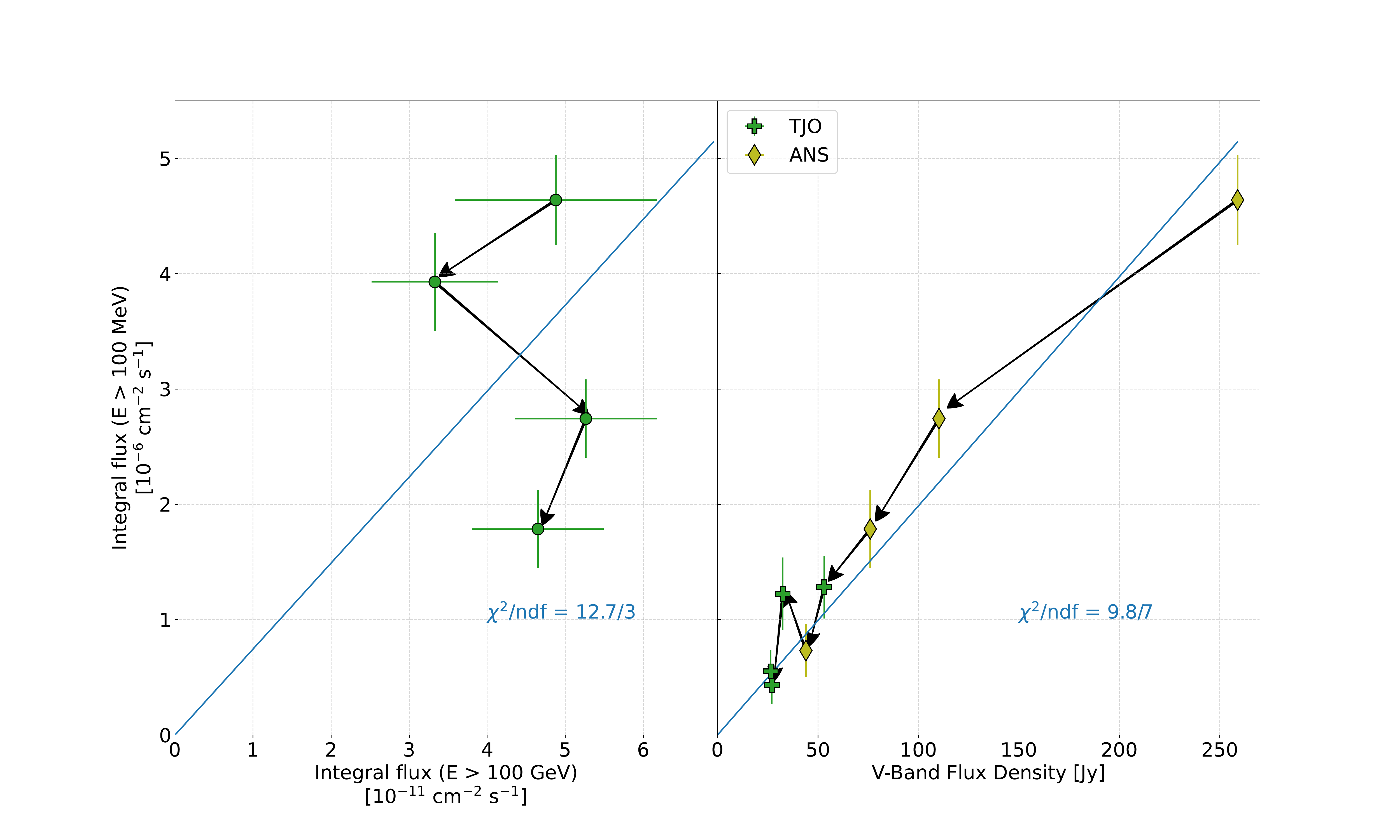}
    %\caption{Left panel: Comparison of the photon flux measured by MAGIC above $100$\,GeV with the one measured by \fermi{} above $100$\,MeV for the first four days of the outburst. Right panel: Comparison of the flux measured by \fermi{} $>100$\,MeV with that of the $V$-band obtained by ANS. Arrows show the sequence of the flux temporal evolution and the blue line shows the linear proportionality fit in both panels. }
    \caption{Comparison of the photon flux measured by \fermi{} above $100$\,MeV with the one measured by MAGIC above $100$\,GeV (left panel) and with that of the $V$-band obtained by ANS (right panel). Arrows show the sequence of the flux temporal evolution and the blue line shows the linear proportionality fit in both panels. Error bars represent 1-sigma statistical uncertainties in the data points.}
    \label{figs:fermi_magic_optical}
\end{figure}

% Extended data Figure 5
\begin{figure}[p]
    \centering
    \includegraphics[trim=30 0 0 0, clip]{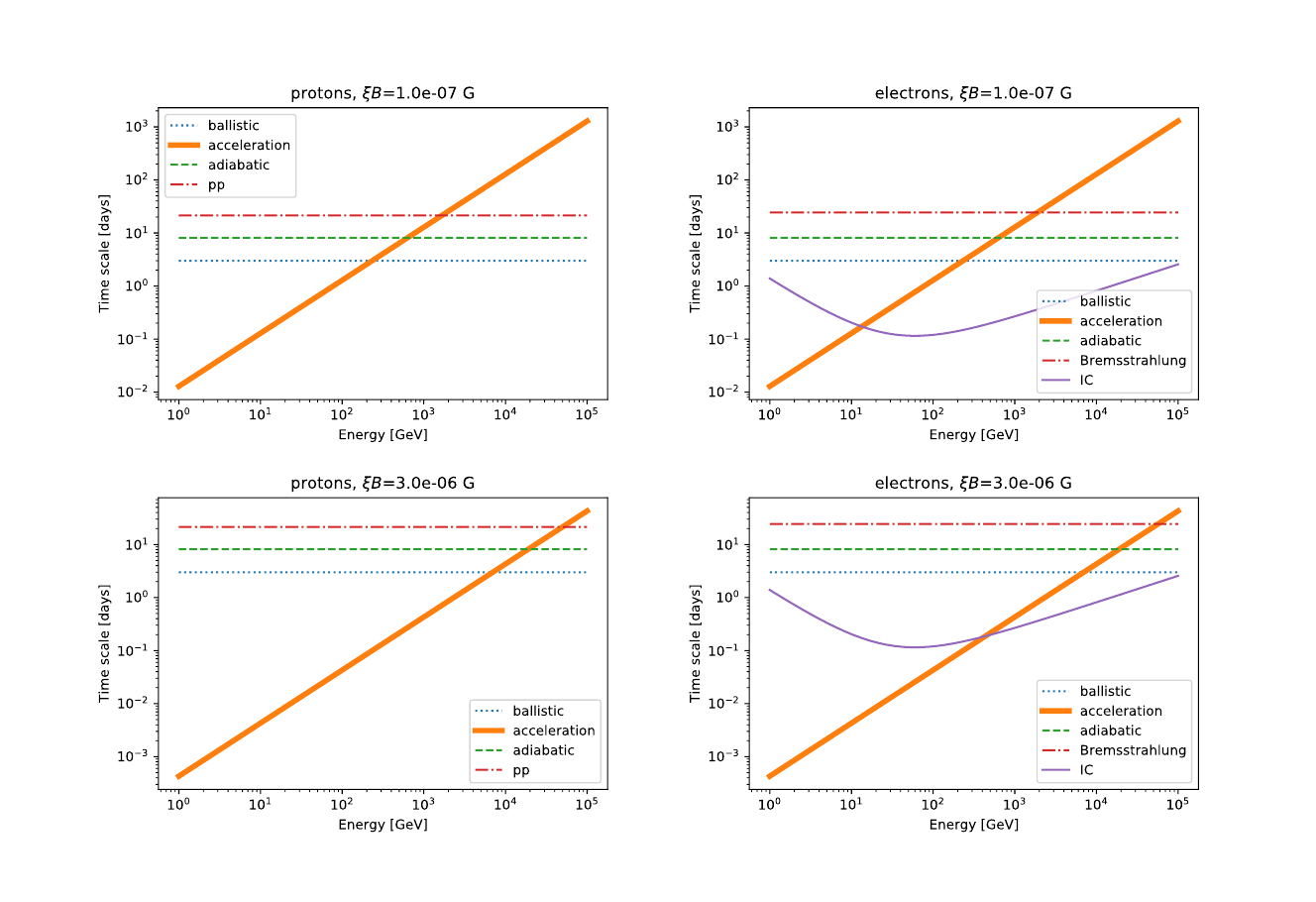}
    \caption{Cooling and acceleration time scale for protons (left panels) and electrons (right panels) for two values of $\xi B$ parameter: $10^{-7}$\,G (top panels) and $3\times10^{-6}$\,G (bottom panels). Assumed parameters (see text for details): $v_{sh} = 4500\,\mathrm{km\,s^{-1}}$, $t = 3$\,d, $R_{ph} = 200\,R_\odot$,  $T_{ph}=8460$\,K, $n_p=6\times10^8\,\mathrm{cm^{-3}}$.}
    \label{figs:timescales}
\end{figure}
%

% Extended data Figure 6
\begin{figure}[p]
    \centering
    \includegraphics[trim=30 0 0 0, clip]{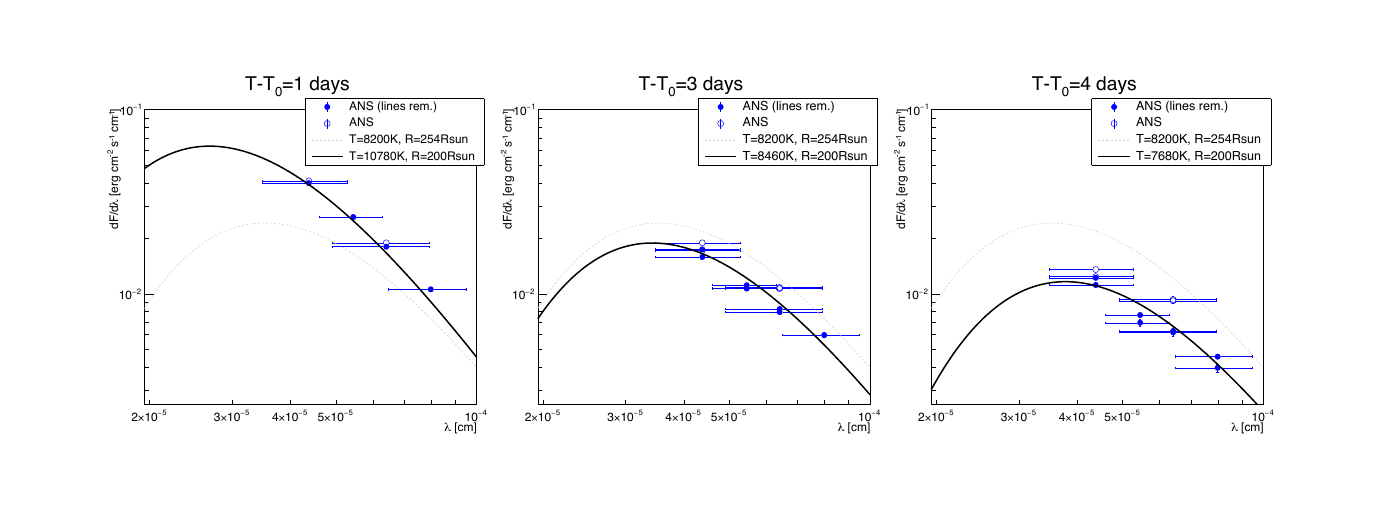}
    \caption{Optical photometry performed by ANS 1, 3, and 4 days \textbf{(see the panel titles)} after the outburst (blue empty markers) corrected for the Galactic absorption. Filled markers show the flux after subtraction of H$\alpha$ and H$\beta$ line contributions. 
    The thick black lines show a black-body emission used in the modeling, while the dashed line shows for comparison the average 2006 spectral fit from \cite{2015NewA...36..128S} (with the photosphere radius corrected to the distance of 2.45\,kpc). Horizontal error bars represent the bandwidth of the filters used.
    }
    \label{figs:spect} 
\end{figure}

% Extended data Figure 7
\begin{figure}[p]
    \centering
    \includegraphics[trim=30 0 0 0, clip]{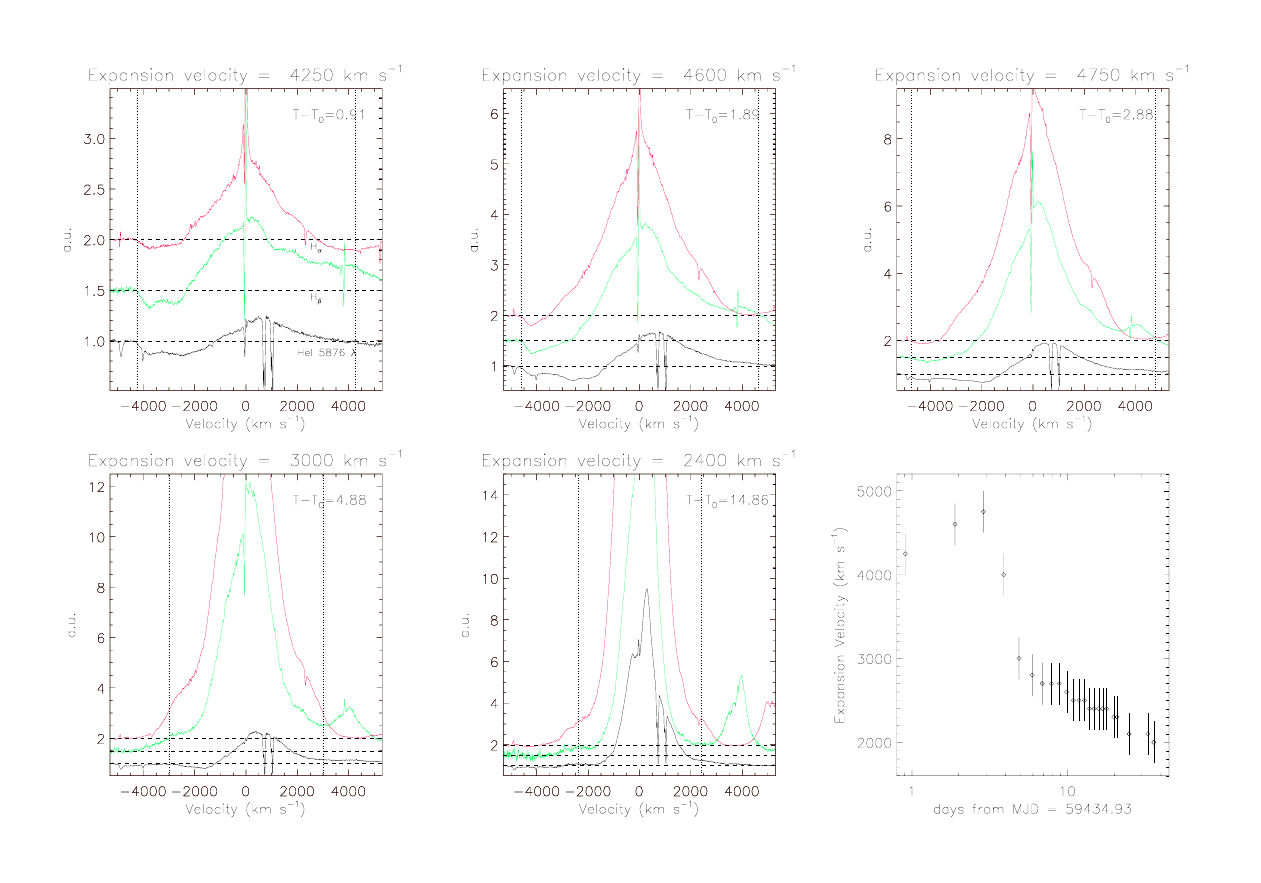}
\caption{Example of H$_\alpha$, H$_\beta$ and He\,I\,5876$\lambda$  P-Cygni profiles used to determine the behavior of the expansion velocity of the expanding envelope (time after the outburst is given in the top right part of each panel). The bottom right panel shows the evolution of the velocity in time. Error bars represent 1-sigma statistical uncertainties in the data points.} \label{figs:line_profiles}
\end{figure}

% Extended data Figure 8
\begin{figure}[p]
    \centering
    \includegraphics[width=0.6\textwidth]{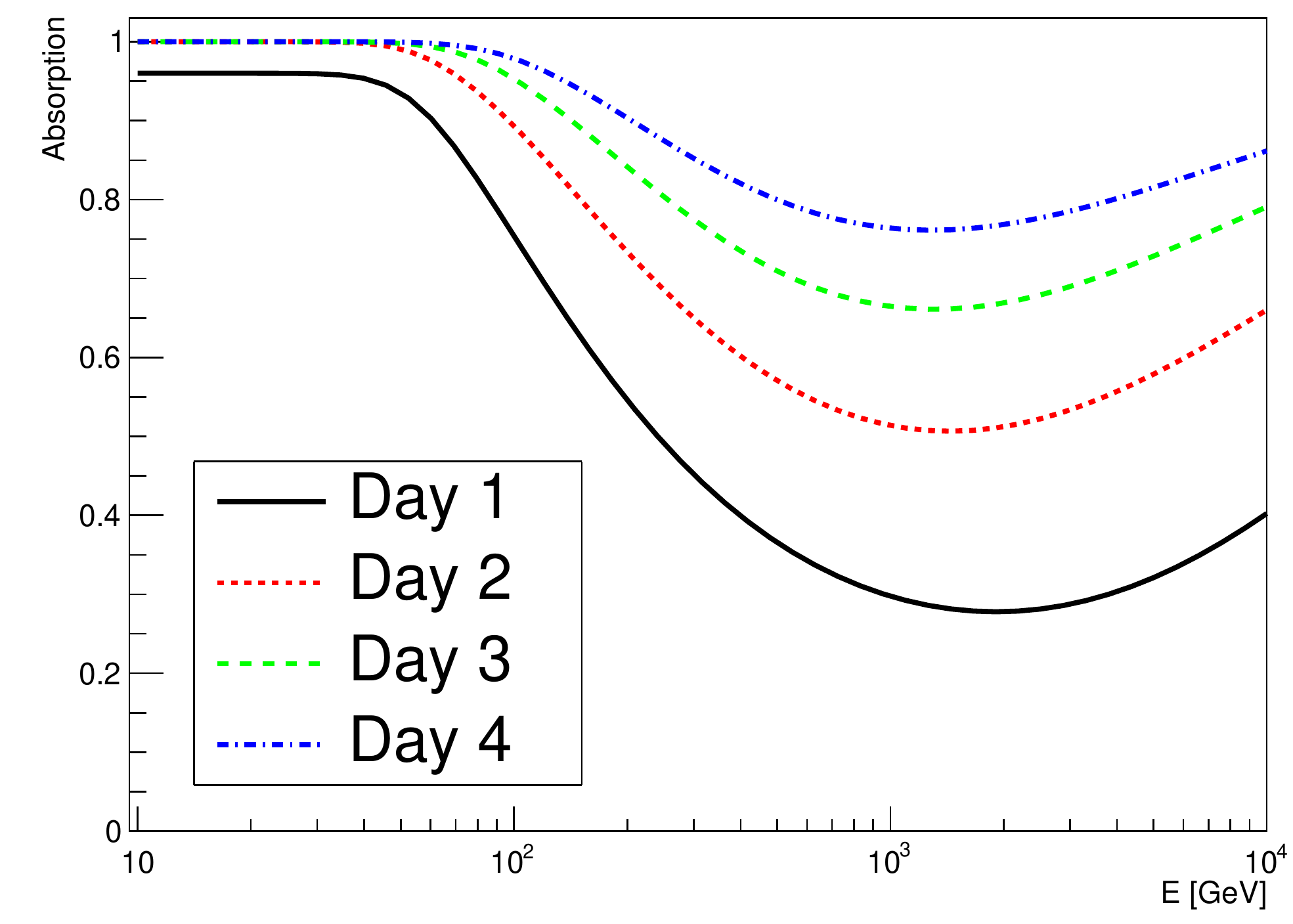}
    \caption{Absorption of the gamma-ray emission on the radiation field of the photosphere and collision with it.
    Assumed parameters: $v_{sh} = 4500\,\mathrm{km\,s^{-1}}$, $R_{ph} = 200\,R_\odot$.
    Temperature of the photosphere is $T_{ph} = 10780$\,K, 9490\,K, 8460\,K and 7680\,K for the time after the nova onset: 1\,d (black solid), 2\,d (red dotted), 3\,d (green dashed), 4\,d (blue dot-dashed) respectively. }
    \label{figs:abs}
\end{figure}

% Extended data Figure 9
\begin{figure}[p]
    \centering
    \includegraphics[width=0.49\textwidth]{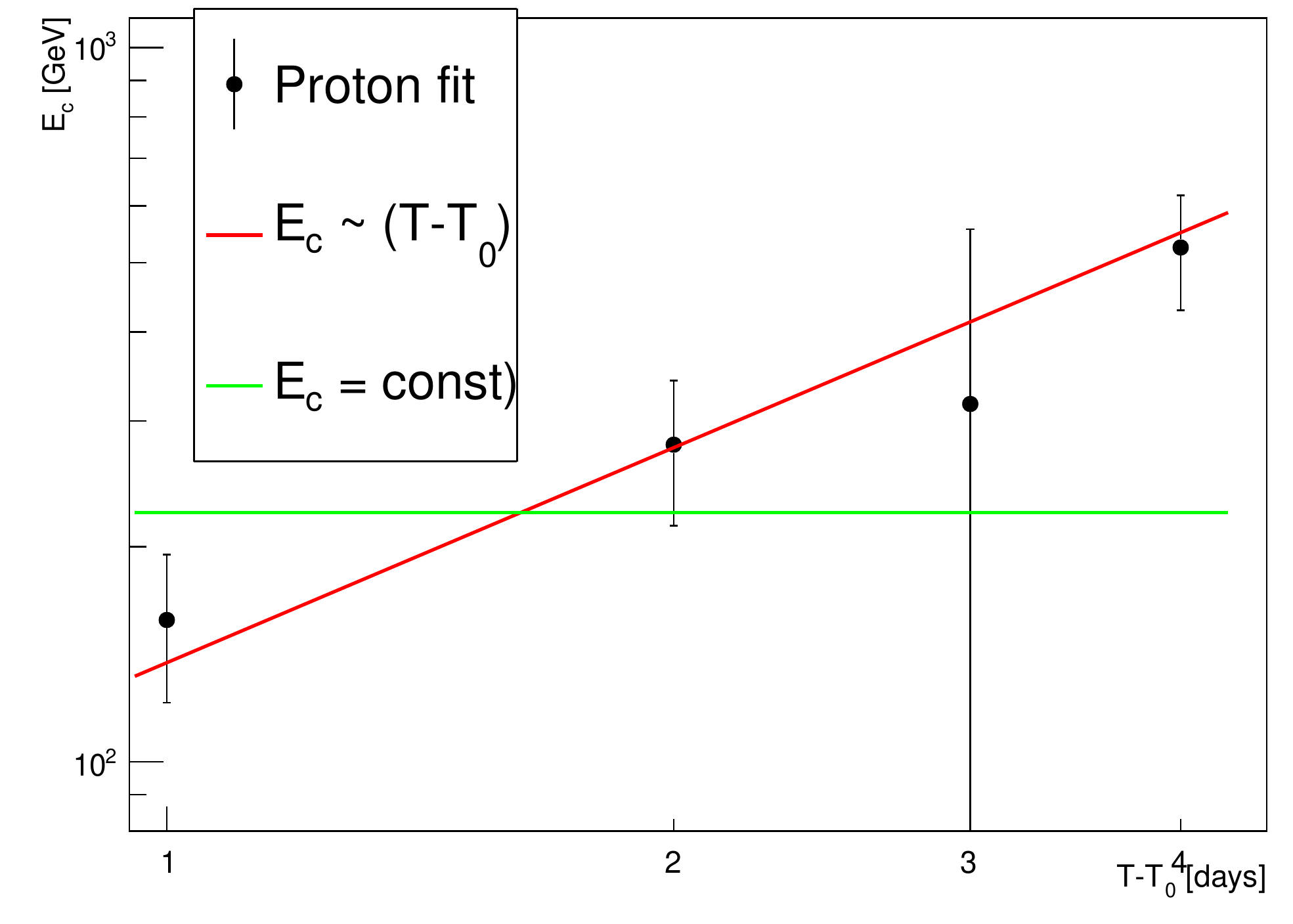}
    \caption{The maximum energy of protons obtained from the theoretical model fits to the daily gamma-ray emission (points) shown in Fig.~\ref{figs:sed_daily}.
    Red and green line show, respectively, the scenario of proportional increase and constant value of maximum energy. Error bars represent 1-sigma statistical uncertainties in the determination of the maximum energy of protons.}
    \label{figs:ecut}
\end{figure}

% Extended data Figure 10
\begin{figure}[p]
    \centering
    \includegraphics[width=0.75\textwidth]{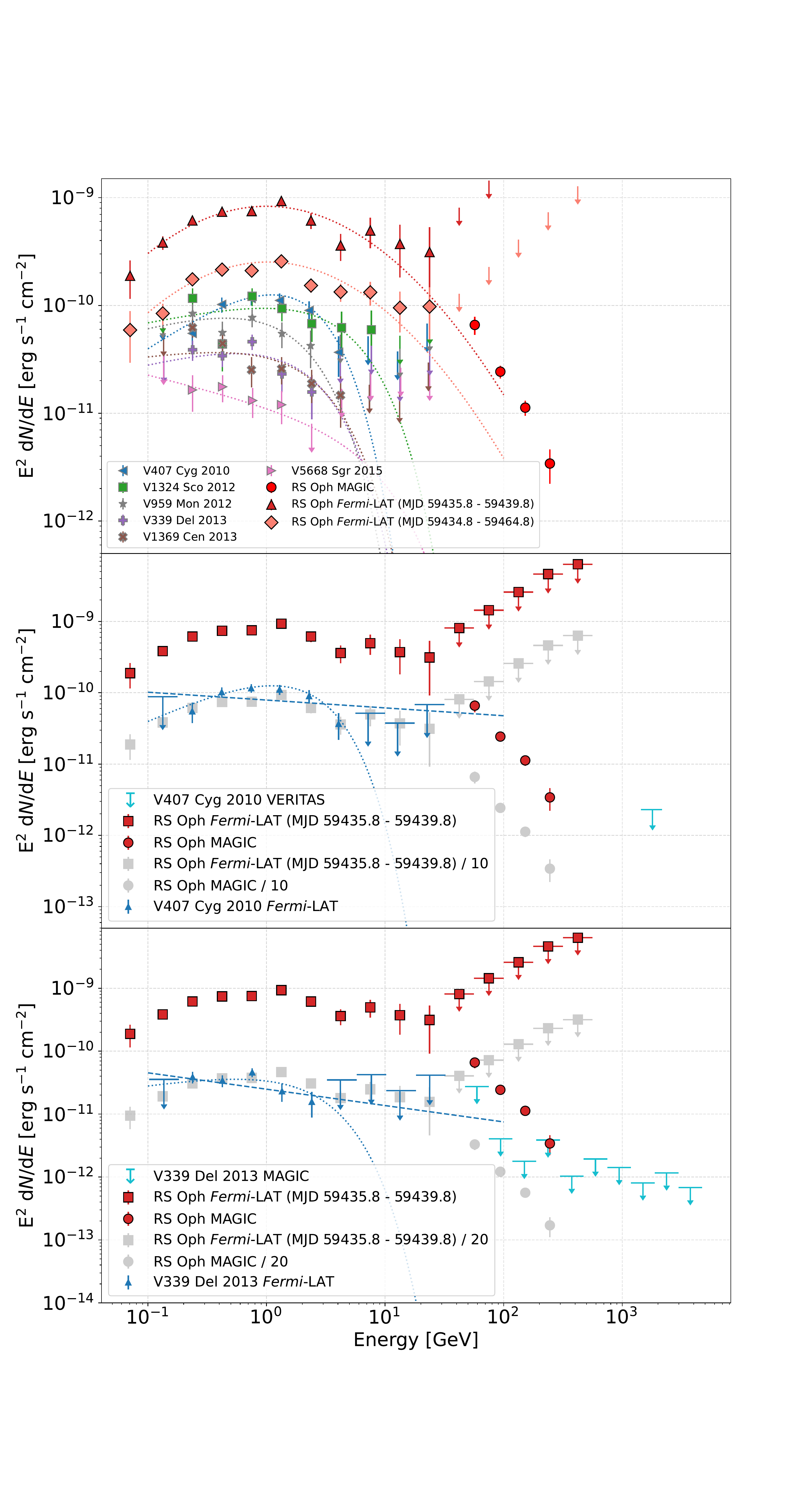}
    \caption{Comparison of \oph{} to other \fermi{}-detected novae. 
    Spectra of other \fermi{}-detected novae are shown in the top panel. 
    Gamma-ray spectra of V407 Cyg (middle panel) and V339 Del (bottom panel) compared to the measured (red) and scaled (gray) \oph{} gamma-ray spectra. Blue triangles and arrows correspond to $Fermi$-LAT measurements and upper limits of V407 Cyg (top) and V339 Del (bottom). Red squares are the $Fermi$-LAT spectrum of \oph{} and red circles the MAGIC one. Gray squares are the $Fermi$-LAT scaled spectrum of \oph{} and gray circles the MAGIC one. Cyan arrows correspond to the VERITAS (V407 Cyg) and MAGIC (V339 Del) upper limits. The dashed blue lines correspond to the best-fit using a single power-law for the $Fermi$-LAT data. The dotted blue lines correspond to the best-fit using a power-law with an exponential cut-off for the $Fermi$-LAT data. 
    %top: \cite{2010Sci...329..817A, 2014Sci...345..554A, 2016ApJ...826..142C}
    % middle and bottom \cite{2010Sci...329..817A, 2014Sci...345..554A,  2015A&A...582A..67A, 2012ApJ...754...77A}
    Data taken from \cite{2010Sci...329..817A, 2014Sci...345..554A, 2016ApJ...826..142C, 2015A&A...582A..67A, 2012ApJ...754...77A}. Error bars represent 1-sigma statistical uncertainties in the data points.}
    \label{figs:v339_and_v407_comparison}
\end{figure}
\clearpage
\renewcommand{\figurename}{Supplementary Figure} 
\setcounter{figure}{0}

\section*{Supplementary material}
\setcounter{page}{1}

\section{The recurrent symbiotic nova \oph{}}\label{sect:oph}

\oph{} is composed by a massive carbon-oxygen white dwarf (WD) \cite{2017ApJ...847...99M} and a M0-2 III mass-donor RG star \cite{1999A&A...344..177A}. 
The orbital solution implies a WD mass of $M_{WD} = 1.2-1.4 M_{\odot}$ and an RG mass of $M_{RG} = 0.68 - 0.80 M_{\odot}$. This nova has shown eight eruptions between 1898 and 2006 \cite{2010ApJS..187..275S}. 
Interestingly, \oph{} was pointed out as a plausible source from which GeV emission can be detected \cite{2007ApJ...663L.101T}. \cite{2008ASPC..401..313H} showed that the prompt hard X/soft gamma-ray emission of the 2006 outburst of \oph{} detected by Swift/BAT \cite{2006ApJ...652..629B} could not be accounted by the decay of radioactive isotopes. \cite{2007ApJ...663L.101T} proved that this emission could be explained via the production of non-thermal particles by diffuse shock acceleration and proposed \oph{} as a possible GeV candidate.
It is also a type Ia Supernova progenitor candidate \cite{2000ApJ...536L..93H,2010ApJS..187..275S}.
The system has a period of $(453.6 \pm 0.4)$ days \cite{2009A&A...497..815B}. 
The system has a circular ($e\approx0$) orbit\cite{1994AJ....108.2259D,2000AJ....119.1375F}, however a mild eccentricity (\mbox{$e = 0.14 \pm 0.03$}) has been claimed as well  \cite{2009A&A...497..815B} probably due to a better coverage of the radial velocity curve. 
The estimation of the wind mass loss rate of the RG in \oph{} is $\sim 5\times 10^{-7} M_\odot\,\mathrm{yr}^{-1}$ \cite{2016MNRAS.457..822B}.
The total matter ejected during nova outburst is difficult to estimate. 
Models of the explosion give values $2\times 10^{-7} - 10^{-6}\,M_\odot$ for different values of the WD mass \cite{2005ApJ...623..398Y}. 
A rough estimation of ``not much more than $10^{-7}\,M_\odot$'' has been given by \cite{2006Natur.442..276S}, however it was based on earlier measurements of the RG wind density that is an order of magnitude lower than the one in \cite{2016MNRAS.457..822B}.  

%\cite{1987ApJ...314..653W}
%The latest outburst happened on MJD $T_0 = 59434.93$. 
An outburst was reported on August 08, 2021 (MJD $T_0 = 59434.93$) \cite{2021ATel14834....1C}. 
The spectroscopy measurements performed in the first few days of the nova show mild acceleration of the ejecta from $3700 - 2700$ km\,s$^{-1}$ at $T_0+0.87$\,d 
to $4200 - 4700$ km\,s$^{-1}$ at $T_0+2$\,d for the H$\alpha$ and H$\beta$ P Cyg lines, respectively \cite{2021ATel14840....1M, 2021ATel14852....1M}.
We assume an ejecta speed  of $v_{sh} = 4500$\,km\,s$^{-1}$, see Methods section~\ref{sect:spectroscopy}. 
The equipartition magnetic field derived from observations starting 20 days after the 2006 outburst is $B = 0.08 - 0.11$\,G  \cite{2008ApJ...688..559R}, while the estimate 18.5\,d after the 1985 nova onset was $0.04$\,G \cite{1985MNRAS.217..205B}. 
In the case of a similar recurrent nova, V1535~Sco, a value of $B$ = 0.13 -- 0.17\,G was measured one week after the outburst \cite{2017ApJ...842...73L} and for V745~Sco $0.03$\,G at the distance of the shock of $4.5\times10^{14}$\,cm \cite{2016MNRAS.456L..49K}.
It should be noted that, as the shock dissipates, $B$ declines over time.

\subsection{Estimates of the \oph{} distance to Earth}
\label{sec:distance}
The distance to \oph{} has been object of intense debate (see \textbf{Supplementary} Table~\ref{tab:dist_table}). 
Historically, a value of $1.6$\,kpc was estimated \cite{1986ApJ...305L..71H, 1987rsop.book.....B} and canonically assumed. 
Dedicated discussions on the distance were performed in the past \cite{2008ASPC..401...52B} in which a distance of $1.4$\,kpc was considered the most likely one. 
The value is however at odds with the mass accumulation rate needed for repetition period of \oph{}. 
Namely, at such assumed distance, the calculated blackbody radius of the secondary star must greatly underfill its Roche lobe \cite{2018MNRAS.481.3033S}.

More recently, the parallax distance to the source of (2.68 $\pm$ 0.16)\,kpc was provided by Gaia \cite{2021A&A...649A...1G}. 
However, as it is discussed in \cite{2018MNRAS.481.3033S}, Gaia Data Release 2 (and Gaia Early Data Release 3) do not have reliable measures of the parallaxes for \oph{}.
The issue arises due to the long-period binary orbit which makes the center of light wobble back and forth with a greater amplitude than the parallax itself, not providing a good fit to the single-star model applied in these data releases.

Using the argument that the RG needs to fill its Roche lobe to efficiently accrete matter onto the WD, the favored distance to the source is $3.1\pm0.5$\,kpc \cite{2008ASPC..401...52B}. \cite{2009ApJ...697..721S} also pointed out that using the light curve information, the most likely distance is $4.3\pm0.7$\,kpc, and the earlier, lower estimates suffered from overestimated absorption along the line of sight (see the discussion in \cite{2018MNRAS.481.3033S}). 
This is however at odds with the expansion velocity of the synchrotron shock, as it was pointed out by \cite{Rupen_2008}, in which they derived a distance of ($2.45\pm0.37$)\,kpc. 
Given all these caveats, in the subsequent calculations we assume the distance to be $2.45$\,kpc.

\section{Multiwavelength view}

In EDF~\ref{figs:fermi_magic_optical}, we compare the integral fluxes (obtained from daily spectral fits) of \fermi{} and MAGIC. 
Fitting the relation with a linear proportionality ($F(>100\,\mathrm{GeV}) \propto F(>0.1\,\mathrm{GeV})$) we obtain $\chi^2/\mathrm{N_{dof}} = 12.7/3$. % 12.72/3 
Therefore simple, achromatic gamma-ray variability is unlikely (chance probability $p = 5.3\times10^{-3}$). 

We also perform a joint fit of \fermi{} and MAGIC data with a Log Parabola function shape:
$dN/dE = f_0 \times (E/E_0)^{-\alpha-\beta \ln(E/E_0)}$ with $E_0 = $130\,GeV.
The fits %for individual nights 
are summarized in \textbf{Supplementary}  Table~\ref{tab:daily_sed_fm}.

In the right panel of EDF~\ref{figs:fermi_magic_optical}, we compare the integral fluxes (obtained from daily spectral fits) of \fermi{} and the differential fluxes for the $V$-band obtained by ANS and TJO. We selected only the coincident data for MJD 59435 - 59445, for which the daily \fermi{} integral fluxes fulfill the condition not to calculate an upper limit. 
We fit a function $y = A\,x$ to the data that gives a $\chi^2/\mathrm{N_{dof}} = 9.8/7$. %9.85/7

Despite the data being consistent with a linear optical-GeV correlation, this does not provide a straight-forward interpretation about the underlying particle population producing this emission. 
As the ejecta propagate away from the WD, the radiation field seen at the shock will in fact decay faster than the observed at Earth optical emission. 
Moreover, as the electrons would cool faster (see Methods section~\ref{sec:model}) than the observed decay of the radiation, we do not expect a linear relation between the IC emission and optical. 
The IC emission would rather follow the rate of injection of particles from the acceleration process. 
This correlation could therefore be related to ejecta or particle acceleration parameters.

\section{Absorption of gamma-ray radiation}
In both electron and proton models, the production of the gamma-ray radiation occurs relatively close to the photosphere $R_{sh}\approx 2.8 (t/\mathrm{[day]}) \times R_{ph}$, a strong thermal source. %1.66
We investigate the effect of absorption of the produced gamma rays on such a radiation field. 
We derive angle-dependent optical depths in vicinity of such thermal source and compute the average absorption as 
\begin{equation}
    \mathrm{Absorption}(E) = 0.5 \int_{0}^\pi e^{-\tau(E, \theta_{ph})}\sin\theta_{ph} \mathrm{d}\theta_{ph},
\end{equation}
where $\theta_{ph}$ is the angle between direction of photon and the radial direction from the center of the photosphere. 
Additionally we assume that photons crossing the photosphere are fully absorbed ($\tau = \infty$). 

Derived absorption of the emission is presented in EDF~\ref{figs:abs} for different days after the nova outburst.
In general the absorption is not very strong, in particular at energies $\lesssim 300$\,GeV, where gamma-ray emission was detected. 
Nevertheless it is taken into account in the modeling. 

\section{Neutrino emission}\label{sect:neutrino}

The gamma-ray emission in hadronic scenario would be accompanied by neutrinos. 
We calculated the neutrino emission corresponding to the proton model presented in Fig.~3 and compared it with limits from the IceCube Collaboration \cite{2021ATel14851....1P}. 
\begin{figure}[t]
    \centering
    \includegraphics[width=0.6\textwidth]{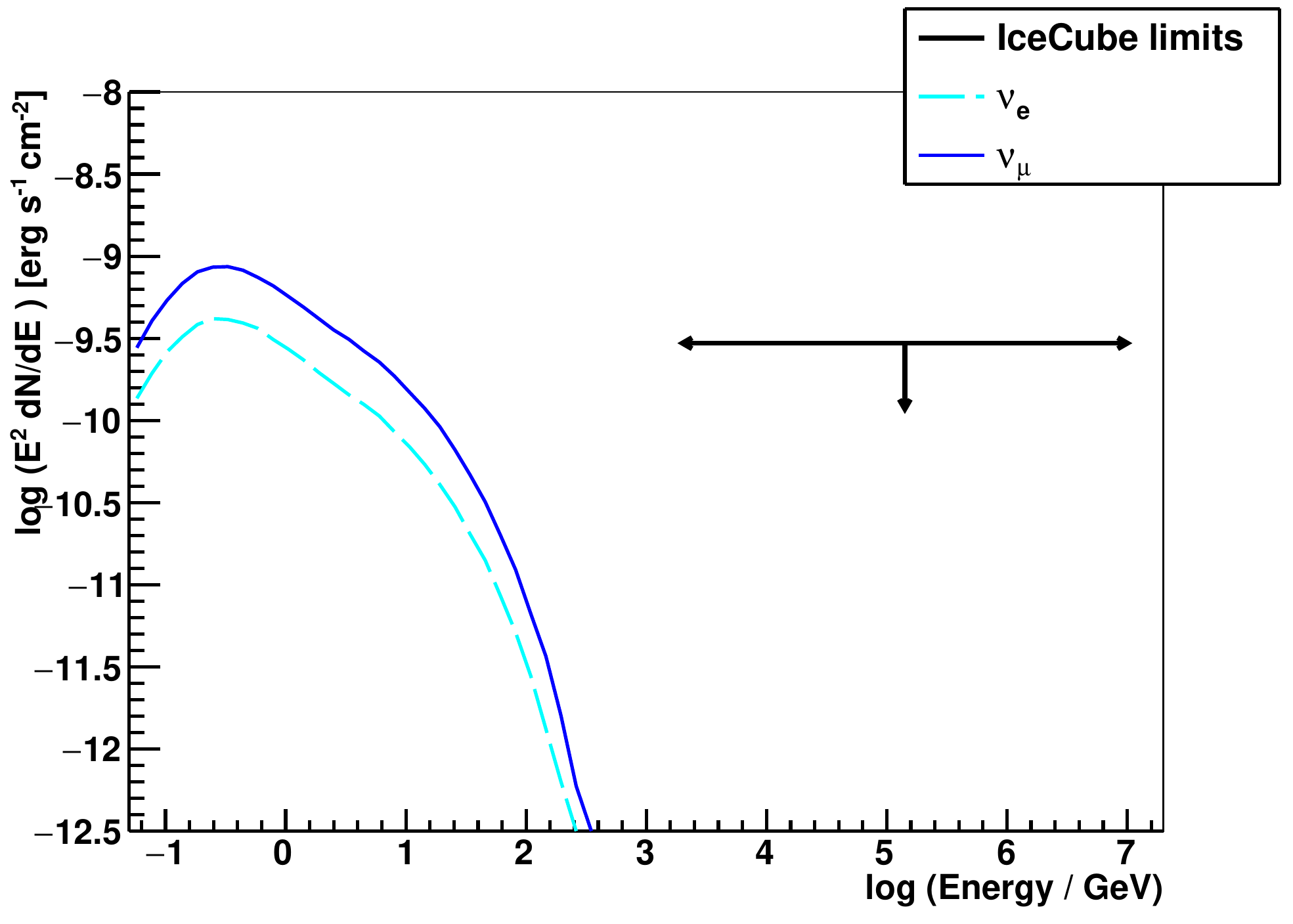}
    \caption{Predicted neutrino emission ($\nu_e$ with cyan dashed line and $\nu_\mu$ with solid blue line) associated to the proton model (see Fig.~\ref{fig3}) compared with 90\% C.L. limits obtained by the IceCube Collaboration \cite{2021ATel14851....1P}.}
    \label{figs:nu}
\end{figure}
It is clear that due to sub-TeV energies achieved by protons, the predicted neutrino emission does not reach energies higher than those of protons and these limits cannot constrain the model (see \textbf{Supplementary} Fig.~\ref{figs:nu}.
We also investigated if SuperKamiokande could have detected neutrino emission associated to the nova outburst.
However, due to low collection area at the GeV energies \cite{the_super_kamiokande_collaboration_2021_4724823} the expected number of events is only of the order of $5\times 10^{-7}$.

\section{Proton-lepton model}

The presence of high-energy protons or electrons is not only dependent on their maximum energies (see Methods section~\ref{sect:maxe}).
Differences in the injection process of electrons and hadrons into the acceleration mechanism (see \cite{2021APh...12702546A} and references therein) can cause preferential dominant acceleration of one or the other type of particles. 
Following \cite{2012PhRvD..86f3011S,2015A&A...582A..67A}, we test as well a model in which both electrons and protons are accelerated in the same shock.
We assume injection with a power-law and exponential cut-off for both particles types. 
The cut-off energies are related by the cooling/acceleration balance (see Methods section~\ref{sect:maxe}). 
The resulting best fit is presented in \textbf{Supplementary} Fig.~\ref{figs:sed_pr_el}.
\begin{figure}[t]
    \centering
    \includegraphics[width=0.6\textwidth]{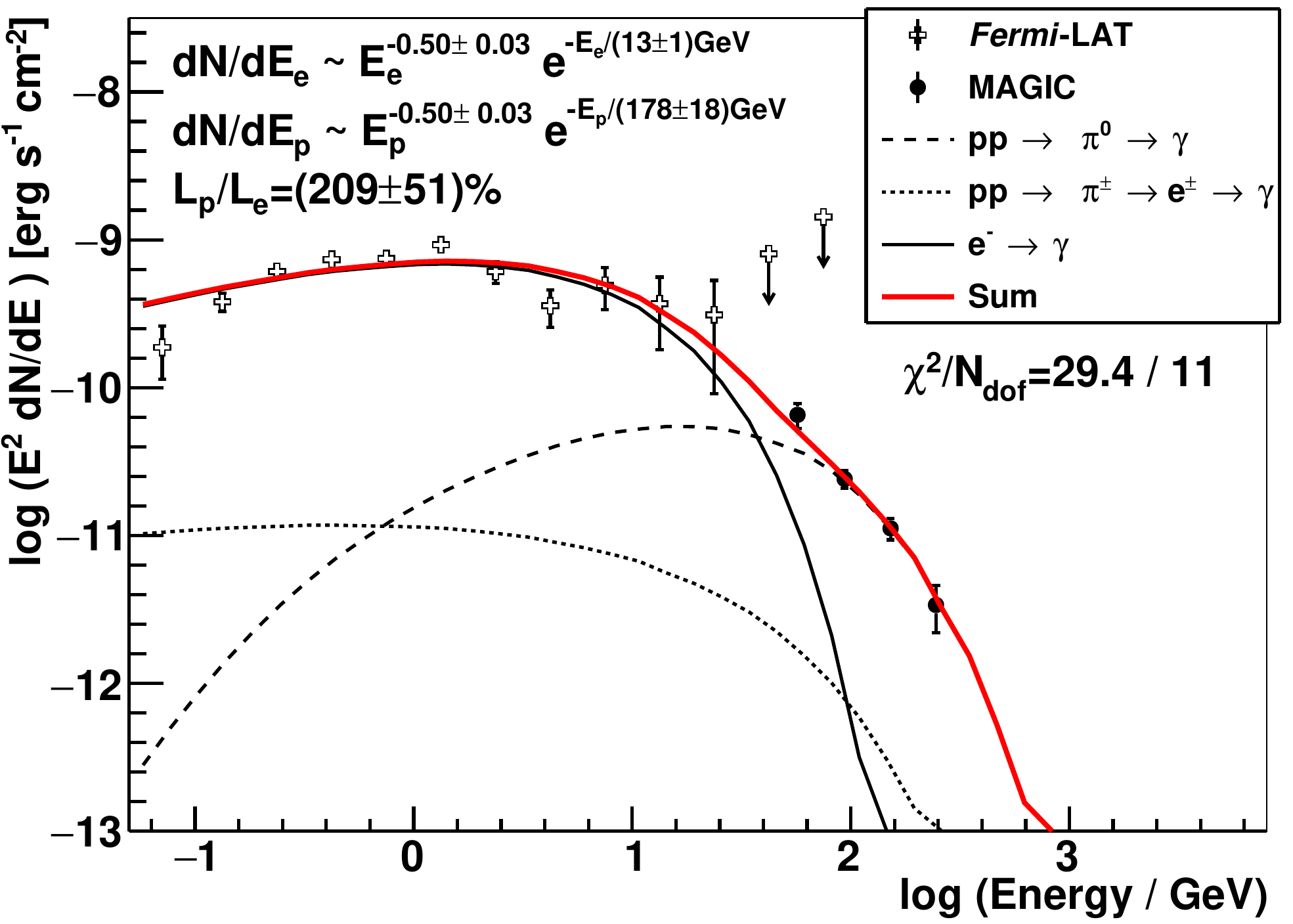}
    \includegraphics[width=0.6\textwidth]{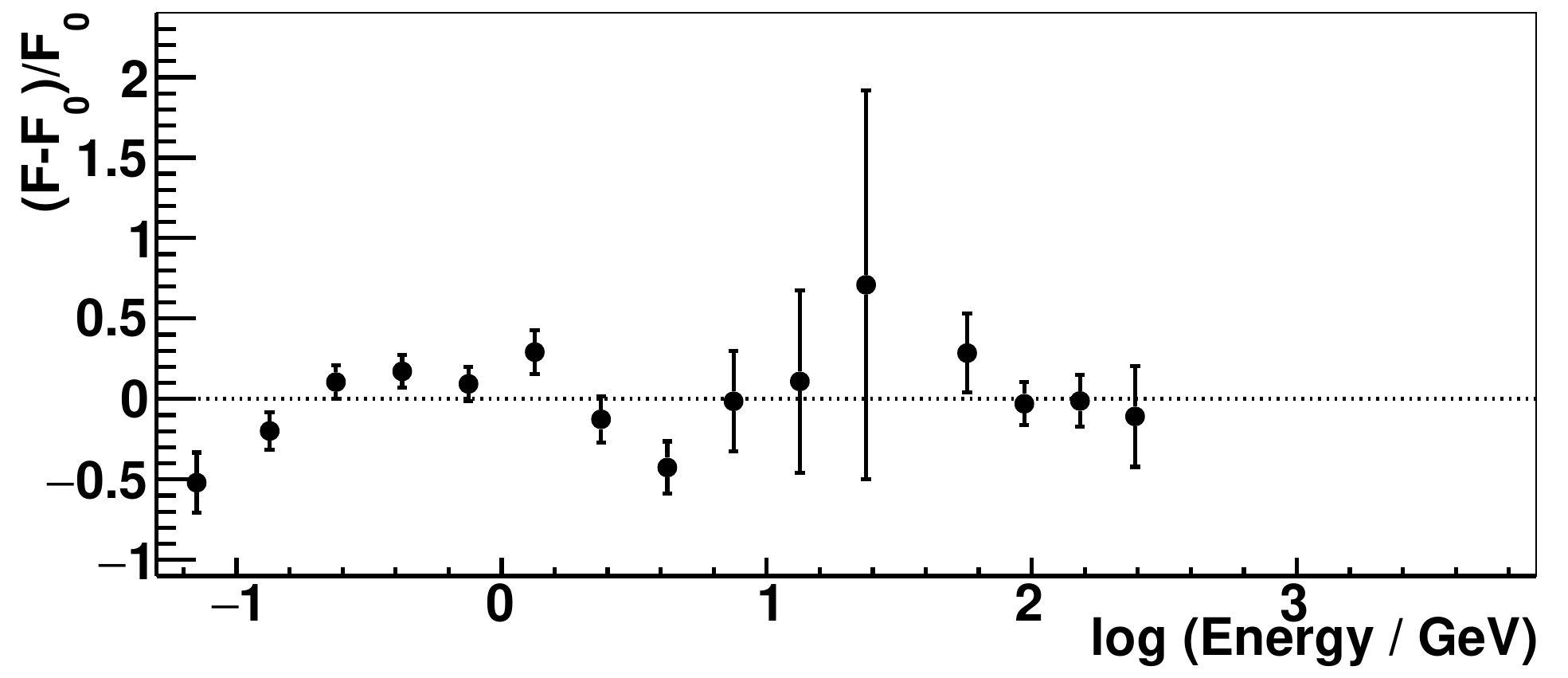}
    \caption{Fit to the \fermi{} and MAGIC SED with a proton-electron model. %%Individual lines as in Fig.~\ref{fig3}.
    The dashed line shows the gamma rays from the $\pi^0$ decay and the dotted line shows the inverse Compton contribution of the secondary $e^\pm$ pairs produced in hadronic interactions.
    $dN/dE_p$ and $dN/dE_e$ report the shape of the proton and electron energy distributions obtained from the fit. 
    The bottom panel shows the fit residuals. Errorbars represent 1-sigma statistical uncertainties in the data points.
}
    \label{figs:sed_pr_el}
\end{figure}
The assumed spectral shape of injected electron and proton populations cannot explain the emission well ($\chi^2/\mathrm{N_{dof}} = 29.4/11$, corresponding to p-value of $2.0\times10^{-3}$). 
The best fit also requires $L_p/L_e\approx 2$, much larger than $\lesssim0.1$ constrained in observations of V337 Del \cite{2015A&A...582A..67A}.

\section{Day-by-day proton modeling}\label{sect:days}

In addition to the modeling of an average state of the source in the first 4 days, we also perform modeling of individual days after the nova onset.
The results of fits with the proton model are shown in left panels of Fig.~\ref{figs:sed_daily}. 
On individual days the preference of the proton model over the electron model is lower, however except for the first night, the electron model provides lower $p$ value.
Summing up over the four days $\chi^2$ increases by $5.2$ despite additional 4 parameters, results in $\Delta$AIC$=13.2$ which corresponds to AIC likelihood ratio of $1.4\times 10^{-3}$.

Interestingly, the spectra show a hint of gradual softening of the power-law component accompanied by and increase of the value of the cut-off energy (see Fig.~\ref{figs:ecut}). 
Such behaviour is in line with the expectations from the cooling and acceleration time scales defined in \textbf{Methods} section \ref{sect:maxe}.
Namely, due to low cooling losses of protons, their maximum energies are mainly determined by the duration of the acceleration.  
The dependence of the maximum energy of protons on time can be fit very well ($\chi^2/\mathrm{N_{dof}} = 0.54/3$) with such a scenario of proportional increase with time (corresponding to $\xi B =$ const). 
Such continuous increase of maximum proton energies could last until the shock is drained up from its energy, or is slowed down by the interstellar medium. 
However, as the target material dilutes with time, the expected gamma-ray emission would fall below the detectability level.  
Constant value of the cut energy can be excluded at chance probability $3.1\times 10^{-3}$ level ($\chi^2/\mathrm{N_{dof}} = 13.9/3$). %13.87/3 
It should be noted that while the fit only considers statistical uncertainties of the reconstructed maximum energy, it is unlikely that any systematic uncertainties would mimick such a gradient as the data are taken over a time span of only a few days in similar observational conditions.

\section{\oph{} in context with other novae}
\label{sec:comparison_novae}
\subsection{Gamma-ray novae}

To put the \oph{} eruption into context, we compared it to other published \fermi{} detections of novae: V407 Cyg 2010 \cite{2010Sci...329..817A}, V3124 Sco 2012, V959 Mon 2012, V339 Del 2013 \cite{2014Sci...345..554A}, V1369 Cen 2013 and V5668 Sgr 2015 \cite{2016ApJ...826..142C}. There are other studies of \fermi{} novae \cite{2020NatAs...4..776A}, apart from several ATels and sub-threshold sources \cite{2018A&A...609A.120F} of classical and symbiotic novae that are not included in the comparison presented in this section. 
It is important to mention that although RS Oph is considered to be a symbiotic nova, it was pointed out \cite{Strope_2010} that even though RNe have nova eruptions on symbiotic stars, they may not share all the properties of symbiotic novae (in particular very slow and low amplitude eruptions without Roche lobe overflow).

On the top panel of  EDF~\ref{figs:v339_and_v407_comparison}, we present a comparison of the \oph{} \fermi{} SED coincident with the MAGIC detection (MJD 59435.8 - 59439.8) and the average of the full flare (MJD 59434.8 - 59464.8) compared to the aforementioned novae. 
We can see that both the flux corresponding to the simultaneous data, and the average flux during the whole eruption are from a factor of a few up to almost two orders of magnitude larger than previously-detected eruptions.

To perform meaningful comparisons, we defined the duration of RS Oph eruption determined by the intervals
spanned by the TS $\geq$ 4 in the daily light
curves \cite{2016ApJ...826..142C}. With this definition, the duration is 30 days, that is comparable to the rest of the \fermi{} published novae. This has not only been the eruption with the highest flux, but also the most luminous one
as it is shown in Fig.~\ref{figs:novae_energy_photons}, for which we have used the results of the fit with an Exponential Cut-Off Power Law fit for E $>$ 0.1 GeV of the average flux from the eruption as in \cite{2016ApJ...826..142C}. 
This statement is dependent on the assumed distance of 2.45\,kpc, and is subject to the uncertainties in the determination of the distances to different novae (see the discussion in  \textbf{Methods} section \ref{sec:distance}).

In \cite{2016ApJ...826..142C} there is the speculation that there is an apparent inverse
relationship of the total energy with gamma-ray durations for  classical novae, that would also roughly be followed by V407 Cyg. The fact that we measured that \oph{} has a factor of a few higher energy emitted in gamma rays, points to intrinsic differences between this eruption and the others detected in classical or symbiotic novae.
The total power of gamma rays emitted from \oph{}, $1.8\times10^{42}$\,erg that is about 
$0.9\times10^{-2} (M_{ej}/(10^{-6}M_{sun})^{-1}(v_{sh}/4500\,\mathrm{km\,s^{-1})^{-2}} (d/2.45\,\mathrm{kpc})^2$ of the kinetic energy of the shock. 

\subsection{Detectability of novae at VHE gamma-ray range}

We perform a comparison of the spectrum of \oph{} eruption with the most similar nova detected at gamma rays so far: V407 Cyg \cite{2010Sci...329..817A}. In the top panel of  EDF~\ref{figs:v339_and_v407_comparison}, we can see the comparison between the average V407 Cyg spectrum measured by \fermi{} during the 22 days of its eruption and the ULs by VERITAS \cite{2012ApJ...754...77A} for a total of $\sim$5 hour observation time, compared to the MAGIC and the \fermi{} flux simultaneous to the MAGIC detection of \oph{}. We also scaled MAGIC and \fermi{} \oph{} flux to reach that of V407 Cyg measured by \fermi{}. We can see that in every case, the UL established by VERITAS on V407 Cyg lies above the extrapolation of the \oph{} flux measured by MAGIC, therefore the \oph{} results are in agreement with the non-detection by VERITAS, assuming that the VHE gamma-ray emission from V407 Cyg follows the same spectral shape as that of \oph{}. 

There are physical differences between classical novae and \oph{}, a recurrent nova with a strong wind from the companion, that could cause the difference in gamma-ray emission. To evaluate the detectability of classical novae, we nevertheless performed a comparison between \oph{} and V339 Del \cite{2015A&A...582A..67A}, observed by MAGIC during its eruption. In the bottom panel of  EDF~\ref{figs:v339_and_v407_comparison}, we can see the average V339 Del spectrum measured by \fermi{} during the 27 days of the eruption and the MAGIC ULs compared to the \oph{} measurement of MAGIC and \fermi{} simultaneous to that of MAGIC. We also scaled the \oph{} flux to reach that of V339 Del measured by \fermi{} for the simultaneous data to MAGIC. We can see that the MAGIC ULs for V339 Del are below the \oph{} measurement, however, if we scale \oph{} flux down, the MAGIC ULs of V339 Del are above the MAGIC measurement of \oph{}. We note the caveat of comparing the average fluxes measured by \fermi{} for previous novae and that simultaneous to MAGIC measurement for \oph{}, in which the state was high.

We can conclude that the detection of \oph{} at VHE gamma rays was possible due to a higher gamma-ray flux, rather than favorable spectral distribution shape, and that other previously detected novae could have emitted photons up to the same energies, that remained undetected due to the sensitivity of the observations. This means that a VHE gamma-ray instrument more sensitive in the $\sim 100$\,GeV energies would open the possibility of detecting a large number of gamma-ray emitting novae if their emission extends up to VHE \cite{2016MNRAS.457.1786M, 2019scta.book.....C}. 

\clearpage
\renewcommand{\tablename}{Supplementary Table} 
\setcounter{table}{0}

% Extended data Table 1
%
\begin{table}%[htbp]
\centering
\caption{Different distances estimated for RS Oph. The distance assumed in subsequent calculations is marked with asterisk.}
\label{tab:dist_table}    
\begin{tabular}{|c c c|}

\hline
Distance [kpc] & Method & Reference\\
\hline
1.6 & H I absorption measurements & \cite{1986ApJ...305L..71H, 1987rsop.book.....B}\\
1.4$^{+0.6}_{-0.2}$ & Several estimations & \cite{2008ASPC..401...52B}\\
2.45 $\pm$ 0.37* & Expansion velocity & \cite{Rupen_2008}\\
 3.1 $\pm$ 0.5 & Requirement of RG filling its Roche lobe & \cite{2008ASPC..401...52B}\\ %\cite{2009ApJ...697..721S}
4.3 $\pm$ 0.7 & Light curve & \cite{2009ApJ...697..721S}\\
2.68 $\pm$ 0.16 & Parallax & {\cite{2018A&A...616A...1G}} \\
\hline
\end{tabular}
\end{table}

% Extended data Table 2

\begin{table}%[htbp]
\centering
\caption{Summary of the MAGIC observation campaign: time of observation slot, observation conditions, total observation time during the slot, effective time after the data selection (only Dark data).}
\label{tab:magic_data}
\begin{tabular}{|c c c c|}

\hline

MJD Start - End & Obs. conditions & Obs. time [h] & Time after cuts [h]  \\
\hline
 %59435.94 - 59435.98 & 1.0 & 1.00 & Dark \\ 
 %59436.89 - 59437.04 & 3.7 & 3.72 & Dark \\ 
 %59437.89 - 59438.03 & 3.3 & 3.27 & Dark \\ 
 %59438.88 - 59439.02 & 3.4 & 3.35 & Dark \\ 
 59435.94 - 59435.98 & Dark & 1.0 & 1.0  \\ %0.96
 59436.89 - 59437.04 & Dark & 3.6 & 3.5  \\ %3.54 
 59437.89 - 59438.03 & Dark & 3.2 & 3.1  \\ %3.09 
 59438.88 - 59439.02 & Dark & 3.2 & 3.2  \\ %3.20
 59439.89 - 59440.02 & Dark + Moon & 3.0 & -  \\ 
 59440.89 - 59441.02 & Dark + Moon & 3.0 & - \\ 
 59444.89 - 59444.91 & Moon & 0.1 & -  \\ 
 59445.88 - 59445.90 & Moon & 0.2 & -  \\ 
%% new period
 59451.89 - 59452.00 & Dark + Moon & 2.1 & 0.5  \\ %0.49
 59452.88 - 59453.01 & Dark + Moon & 2.9 & 1.6  \\ %1.62 
 59453.88 - 59454.00 & Dark + Moon & 2.7 & 2.0  \\ %2.05 
 59454.87 - 59454.98 & Dark & 2.5 & -  \\ 
 59455.87 - 59455.97 & Dark & 2.3 & 2.3  \\ %2.28
 59456.87 - 59456.97 & Dark & 2.3 & 2.3  \\ %2.28 
 59458.89 - 59458.97 & Dark & 1.9 & 1.9 \\ %1.89 

 \hline
\end{tabular}
\end{table}

% Extended data Table 3

%

\begin{table}%[htbp]
\centering
\caption{Spectral fit results (normalization $f_0$ at $E_0$ = 130\,GeV, and photon index $\alpha$) of MAGIC data: daily and the combined emission from the first four days. Errors represent 1-sigma statistical uncertainties in the fits.}
\label{tab:daily_sed}
\begin{tabular}{|c c c c c|}

\hline
MJD & $f_0$ [$10^{-10}$ TeV$^{-1}$cm$^{-2}$s$^{-1}$] & $\alpha$ &  $\chi^2/\mathrm{N_{dof}}$ \\ 
\hline
%
% 59435.94 - 59435.98 & $2.88^{+0.82}_{-1.01}$ & 3.92$^{+0.51}_{-0.68}$ & 150\\
% 59436.89 - 59437.04 & $1.90^{+0.56}_{-0.55}$ & 4.71$^{+0.34}_{-0.42}$ & 150\\
% 59437.89 - 59438.03 & $2.97^{+0.50}_{-0.51}$ & 3.70$^{+0.28}_{-0.32}$ & 150\\
% 59438.88 - 59439.02 & $2.82^{+0.48}_{-0.49}$ & 3.78$^{+0.25}_{-0.28}$ & 150\\ \hline
% 59435.94 - 59439.02 & $2.60^{+0.30}_{-0.31}$ & 4.07$^{+0.18}_{-0.20}$ & 150\\
% updated 20210921 to E0 = 130 GeV  
 59435.94 - 59435.98 & $5.0^{+1.3}_{-1.5}$ & 3.92$^{+0.51}_{-0.68}$ & 5.6/5 \\ %5.65
 59436.89 - 59437.04 & $3.73^{+0.92}_{-0.94}$ & 4.71$^{+0.34}_{-0.42}$ & 5.1/5 \\ %5.12
 59437.89 - 59438.03 & $5.03^{+0.81}_{-0.80}$ & 3.70$^{+0.28}_{-0.32}$ & 3.6/5 \\ %3.61
 59438.88 - 59439.02 & $4.83^{+0.77}_{-0.77}$ & 3.78$^{+0.25}_{-0.28}$ & 10.3/5 \\ %10.28
 \hline
 59435.94 - 59439.02 & $4.66^{+0.47}_{-0.48}$ & 4.07$^{+0.18}_{-0.20}$ & 5.9/5 \\ %5.89
\hline
\end{tabular}
\end{table}

% Extended data Table 4

%

\begin{table}%[htbp]
\centering
\caption{\fermi{} 1-Day Average Integral Flux. Values with $^{*}$ are fixed in the UL calculation. Errors represent 1-sigma statistical uncertainties in the fits.}
\label{tab:lat_daily_flux}
\begin{tabular}{|c c c c c|}

\hline
MJD Start - End & TS & \begin{tabular}{@{}c@{}}Integral Flux ($>$ 0.1 GeV) \\ (10$^{-7}$ photons\,cm$^{-2}$\,s$^{-1}$)\end{tabular} & $\alpha$ & $\beta$ \\
\hline

59431.45 - 59432.45 & 0.4 & $<11.0$ & $2.0^{*}$ & $0.0^{*}$\\
59432.45 - 59433.45 & 0.0 & $<10.8$ & $2.0^{*}$ & $0.0^{*}$\\ 
59433.45 - 59434.45 & 0.0 & $<10.7$ & $2.0^{*}$ & $0.0^{*}$\\ 
59434.45 - 59435.45 & 191.2 & 18.8 $\pm$ 3.1 & 2.16 $\pm$ 0.13 & 0.054 $\pm$ 0.076\\
59435.45 - 59436.45 & 1006.9 & 46.4 $\pm$ 3.9 & 1.96 $\pm$ 0.078 & 0.197 $\pm$ 0.051\\ 
59436.45 - 59437.45 & 501.0 & 39.3 $\pm$ 4.3 & 2.123 $\pm$ 0.099 & 0.175 $\pm$ 0.066\\ 
59437.45 - 59438.45 & 433.4 & 27.4 $\pm$ 3.4 & 1.955 $\pm$ 0.095 & 0.169 $\pm$ 0.065\\ 
59438.45 - 59439.45 & 197.8 & 17.9 $\pm$ 3.4 & 2.12 $\pm$ 0.16 & 0.24 $\pm$ 0.12\\ 
59439.45 - 59440.45 & 172.3 & 12.8 $\pm$ 2.7& 1.96 $\pm$ 0.16 & 0.22 $\pm$ 0.11\\ 
59440.45 - 59441.45 & 94.6 & 7.3 $\pm$ 2.3 & 1.63 $\pm$ 0.21 & 0.16 $\pm$ 0.11\\
59441.45 - 59442.45 & 97.6 & 12.2 $\pm$ 3.2 & 1.99 $\pm$ 0.17 & 0.15 $\pm$ 0.12\\
59442.45 - 59443.45 & 60.3 & 4.3 $\pm$ 1.6 & 1.96 $\pm$ 0.42 & 0.95 $\pm$ 0.58\\
59443.45 - 59444.45 & 82.0 & 5.5 $\pm$ 1.9 & 1.58 $\pm$ 0.28 & 0.39 $\pm$ 0.21\\ 
59444.45 - 59445.45 & 2.5 & $<11.4$ & $2.0^{*}$ & $0.0^{*}$\\
59445.45 - 59446.45 & 24.9 & 10.0 $\pm$ 4.1 & 2.65 $\pm$ 0.61 & 0.18 $\pm$ 0.33\\
59446.45 - 59447.45 & 16.0 & $<11.9$ & $2.0^{*}$ & $0.0^{*}$\\ 
59447.45 - 59448.45 & 17.1 & $<11.7$ & $2.0^{*}$ & $0.0^{*}$\\
59448.45 - 59449.45 & 28.5 & $<12.1$ & $2.0^{*}$ & $0.0^{*}$\\ 
59449.45 - 59450.45 & 3.2 & $<11.5$ & $2.0^{*}$ & $0.0^{*}$\\
59450.45 - 59451.45 & 27.6 & 4.0 $\pm$ 2.3 & 1.93 $\pm$ 0.34 & 0.16 $\pm$ 0.22\\
59451.45 - 59452.45 & 0.7 & $<11.2$ & $2.0^{*}$ & $0.0^{*}$\\
59452.45 - 59453.45 & 2.4 & $<11.1$ & $2.0^{*}$ & $0.0^{*}$\\ 
59453.45 - 59454.45 & 13.3 & $<11.3$ & $2.0^{*}$ & $0.0^{*}$\\
59454.45 - 59455.45 & 20.2 & 6.1 $\pm$ 3.1 & 2.48 $\pm$ 0.49 & 0.15 $\pm$ 0.28\\
59455.45 - 59456.45 & 5.0 & $<11.6$ & $2.0^{*}$ & $0.0^{*}$\\
59456.45 - 59457.45 & 7.5 & $<11.7$ & $2.0^{*}$ & $0.0^{*}$\\ 
59457.45 - 59458.45 & 1.0 & $<11.0$ & $2.0^{*}$ & $0.0^{*}$\\
59458.45 - 59459.45 & 5.8 & $<11.7$ & $2.0^{*}$ & $0.0^{*}$\\ 
59459.45 - 59460.45 & 6.1 & $<11.6$ & $2.0^{*}$ & $0.0^{*}$\\ 
59460.45 - 59461.45 & 12.8 & 4.2 $\pm$ 2.3 & 2.26 $\pm$ 0.46 & 0.51 $\pm$ 0.26  \\
59461.45 - 59462.45 & 8.3 & $<11.3$ & $2.0^{*}$ & $0.0^{*}$ \\
59462.45 - 59463.45 & 4.4 & $<11.4$ & $2.0^{*}$ & $0.0^{*}$ \\
59463.45 - 59464.45 & 4.6 & $<11.5$ & $2.0^{*}$ & $0.0^{*}$ \\
59464.45 - 59465.45 & 1.9 & $<10.8$ & $2.0^{*}$ & $0.0^{*}$\\

 \hline
\end{tabular}
\end{table}

% Extended data Table 5

%

\begin{table}%[htbp]
\centering
\caption{\fermi{} 3-Day Average Integral Flux. Values with $^{*}$ are fixed in the UL calculation. Errors represent 1-sigma statistical uncertainties in the fits.}
\label{tab:lat_3daily_flux}
\begin{tabular}{|c c c c c|}

\hline
MJD Start - End & TS & \begin{tabular}{@{}c@{}}Integral Flux ($>$ 0.1 GeV) \\ (10$^{-7}$ photons\,cm$^{-2}$\,s$^{-1}$)\end{tabular} & $\alpha$ & $\beta$ \\
\hline

59431.45 - 59434.45 & 0.0 & $<10.5$ & $2.0^{*}$ & $0.0^{*}$ \\ 
59434.45 - 59437.45 & 1621.7 & 34.0 $\pm$ 2.1 & 2.041 $\pm$ 0.053 & 0.160$\pm$ 0.040\\
59437.45 - 59440.45 & 797.5 & 19.7 $\pm$ 1.8 & 1.999 $\pm$ 0.072 & 0.200$\pm$ 0.051\\
59440.45 - 59443.45 & 244.0 & 8.2 $\pm$ 1.5 & 1.83 $\pm$ 0.12 & 0.200$\pm$ 0.084\\
59443.45 - 59446.45 & 80.6 & 4.0 $\pm$ 1.3 & 1.90 $\pm$ 0.22 & 0.28$\pm$ 0.16\\
59446.45 - 59449.45 & 49.0 & 1.62 $\pm$ 0.67 & 1.47 $\pm$ 0.40 & 0.55$\pm$ 0.32\\
59449.45 - 59452.45 & 25.4 & 2.3 $\pm$ 1.3 & 2.01 $\pm$ 0.32 & 0.16$\pm$ 0.20\\
59452.45 - 59455.45 & 27.4 & 2.0 $\pm$ 1.2 & 1.93 $\pm$ 0.29 & 0.03$\pm$ 0.13\\
59455.45 - 59458.45 & 11.3 & $<10.8$  & $2.0^{*}$ & $0.0^{*}$\\
59458.45 - 59461.45 & 10.8 & $<11.6$  & $2.0^{*}$ & $0.0^{*}$\\
59461.45 - 59464.45 & 9.0 & $<11.2$ & $2.0^{*}$ & $0.0^{*}$\\

 \hline
\end{tabular}
\end{table}

% Extended data Table 6

%

\begin{table}%[htbp]

\centering
\caption{The observed optical magnitude of \oph{}. Errors represent 1-sigma statistical uncertainties in the measurements.}
\label{tab:magnitudes}
\begin{tabular}{|c c c c c c|}
\hline
MJD & Telescope & $B$ & $V$ & $R_c$ & $I_c$ \\  
\hline
59435.913  &  ANS  &  5.667 $\pm$ 0.011  &  4.884 $\pm$ 0.014  &  4.194 $\pm$ 0.016  &  3.544 $\pm$ 0.018  \\
59437.824  &  ANS  &  6.611 $\pm$ 0.011  &  5.855 $\pm$ 0.014  &  4.816 $\pm$ 0.016  &  4.164 $\pm$ 0.020  \\
59437.850  &  ANS  &  6.508 $\pm$ 0.013  &  5.811 $\pm$ 0.022  &  4.801 $\pm$ 0.028  &  4.167 $\pm$ 0.032  \\
59438.820  &  ANS  &  6.968 $\pm$ 0.034  &  6.321 $\pm$ 0.045  &  4.966 $\pm$ 0.046  &  4.611 $\pm$ 0.054  \\
59438.845  &  ANS  &  6.871 $\pm$ 0.012  &  6.215 $\pm$ 0.024  &  4.982 $\pm$ 0.036  &  4.459 $\pm$ 0.030  \\
59439.812  &  ANS  &  7.288 $\pm$ 0.014  &  6.637 $\pm$ 0.017  &  5.270 $\pm$ 0.019  &  4.856 $\pm$ 0.020  \\
59439.881  &  TJO  &  7.303 $\pm$ 0.012  &  6.605 $\pm$ 0.011  &  -    &  4.850 $\pm$ 0.015  \\
59440.817  &  TJO  &  -  &  6.895 $\pm$ 0.015  &  5.482 $\pm$ 0.014  &  5.103 $\pm$ 0.011  \\
59440.818  &  ANS  &  7.555 $\pm$ 0.008  &  6.923 $\pm$ 0.009  &  5.421 $\pm$ 0.011  &  5.050 $\pm$ 0.014  \\
59440.840  &  ANS  &  7.445 $\pm$ 0.008  &  6.808 $\pm$ 0.014  &  5.285 $\pm$ 0.026  &  4.911 $\pm$ 0.021  \\
59441.817  &  TJO  &  -   &  7.137 $\pm$ 0.020  &  5.587 $\pm$ 0.014  &  5.291 $\pm$ 0.013  \\
59442.886  &  TJO  &  -   &  7.328 $\pm$ 0.011  &  -   &   -  \\
59442.936  &  TJO  &  7.974 $\pm$ 0.012  &  7.335 $\pm$ 0.011  &  5.704 $\pm$ 0.011  &  5.431 $\pm$ 0.011  \\
59443.824  &  ANS  &  8.053 $\pm$ 0.008  &  7.436 $\pm$ 0.010  &  5.745 $\pm$ 0.011  &  5.537 $\pm$ 0.014  \\
59443.886  &  TJO  &  8.014 $\pm$ 0.012  &  7.359 $\pm$ 0.011  &  5.740 $\pm$ 0.011  &  5.537 $\pm$ 0.011  \\
59444.313  &  ANS  &  8.129 $\pm$ 0.008  &  7.508 $\pm$ 0.009  &  5.797 $\pm$ 0.011  &  5.618 $\pm$ 0.014  \\
59444.926  &  TJO  &  8.184 $\pm$ 0.013  &  7.616 $\pm$ 0.011  &  5.881 $\pm$ 0.011  &  5.719 $\pm$ 0.011  \\
59445.849  &  TJO  &  -   &  7.688 $\pm$ 0.012  &  5.967 $\pm$ 0.011  &  5.782 $\pm$ 0.011  \\
59446.826  &  ANS  &  8.367 $\pm$ 0.010  &  7.752 $\pm$ 0.012  &  6.003 $\pm$ 0.014  &  5.875 $\pm$ 0.016  \\
59446.897  &  TJO  &  8.381 $\pm$ 0.013  &  7.695 $\pm$ 0.011  &  6.011 $\pm$ 0.011  &  5.867 $\pm$ 0.011  \\
59447.807  &  ANS  &  8.582 $\pm$ 0.008  &  7.893 $\pm$ 0.010  &  6.115 $\pm$ 0.012  &  6.037 $\pm$ 0.014  \\
59447.917  &  TJO  &  8.557 $\pm$ 0.014  &  7.914 $\pm$ 0.012  &  6.105 $\pm$ 0.011  &  6.024 $\pm$ 0.011  \\
59449.808  &  ANS  &  8.655 $\pm$ 0.008  &  7.961 $\pm$ 0.014  &  6.157 $\pm$ 0.016  &  6.139 $\pm$ 0.014  \\
59449.899  &  TJO  &  8.719 $\pm$ 0.028  &  8.193 $\pm$ 0.019  &  6.255 $\pm$ 0.013  &  -   \\
59452.824  &  ANS  &  8.993 $\pm$ 0.007  &  8.431 $\pm$ 0.009  &  6.374 $\pm$ 0.014  &  6.492 $\pm$ 0.015  \\
59454.882  &  TJO  &  9.027 $\pm$ 0.012  &  8.433 $\pm$ 0.011  &  -   &   -  \\
59456.813  &  ANS  &  9.106 $\pm$ 0.008  &  8.532 $\pm$ 0.016  &  6.645 $\pm$ 0.017  &  6.728 $\pm$ 0.018  \\
59458.804  &  ANS  &  9.181 $\pm$ 0.016  &  8.594 $\pm$ 0.019  &  6.535 $\pm$ 0.048  &  6.774 $\pm$ 0.038  \\
59459.801  &  ANS  &  9.277 $\pm$ 0.009  &  8.748 $\pm$ 0.010  &  6.792 $\pm$ 0.011  &  6.984 $\pm$ 0.012  \\
59459.805  &  ANS  &  9.305 $\pm$ 0.016  &  8.695 $\pm$ 0.018  &  6.641 $\pm$ 0.051  &  6.899 $\pm$ 0.041  \\
59459.834  &  ANS  &  9.385 $\pm$ 0.007  &  8.879 $\pm$ 0.008  &  6.862 $\pm$ 0.009  &  7.043 $\pm$ 0.014  \\
59460.801  &  ANS  &  9.403 $\pm$ 0.009  &  8.848 $\pm$ 0.015  &  6.894 $\pm$ 0.016  &  7.086 $\pm$ 0.017  \\
59460.808  &  ANS  &  9.461 $\pm$ 0.007  &  8.958 $\pm$ 0.008  &  6.913 $\pm$ 0.010  &  7.148 $\pm$ 0.014  \\
59460.871  &  TJO  &  9.361 $\pm$ 0.013  &  8.918 $\pm$ 0.011  &  6.972 $\pm$ 0.011  &  7.134 $\pm$ 0.011  \\
59461.825  &  ANS  &  9.597 $\pm$ 0.007  &  8.997 $\pm$ 0.014  &  7.039 $\pm$ 0.015  &  7.230 $\pm$ 0.016  \\
59462.796  &  ANS  &  9.690 $\pm$ 0.033  &  9.070 $\pm$ 0.043  &  6.995 $\pm$ 0.070  &  7.228 $\pm$ 0.054  \\
59463.805  &  ANS  &  9.627 $\pm$ 0.008  &  9.087 $\pm$ 0.011  &  7.095 $\pm$ 0.012  &  7.307 $\pm$ 0.012  \\
59463.883  &  ANS  &  9.789 $\pm$ 0.010  &  9.296 $\pm$ 0.018  &  7.299 $\pm$ 0.021  &  7.459 $\pm$ 0.024  \\
59465.791  &  ANS  &  9.781 $\pm$ 0.017  &  9.161 $\pm$ 0.021  &  7.089 $\pm$ 0.034  &  7.363 $\pm$ 0.039  \\
\hline
\end{tabular}
\end{table}

% Extended data Table 7

%
\begin{table}%[htbp]

\centering
\caption{The observed optical magnitude ($B$ and $R_c$ band) of RS Oph after removing the contribution of $H_{\alpha}$ and $H_{\beta}$ emission lines. Errors represent 1-sigma statistical uncertainties in the measurements.}
\label{tab:corr_mag}
\begin{tabular}{|c c c c c c|}
\hline
MJD & Telescope & ${H_{\beta}/B}$ (\%) &  $B$ & ${H_{\alpha}/R_c}$ (\%) & $R_c$ \\  
\hline
59435.913  &  ANS  &  3   &  5.698 $\pm$ 0.011  &  5   &  4.247 $\pm$ 0.016  \\
59437.824  &  ANS  &  9   &  6.701 $\pm$ 0.011  &  31  &  5.134 $\pm$ 0.016  \\
59437.850  &  ANS  &  9   &  6.598 $\pm$ 0.013  &  34  &  5.093 $\pm$ 0.028  \\
59438.820  &  ANS  &  11  &  7.081 $\pm$ 0.034  &  48  &  5.393 $\pm$ 0.046  \\
59438.845  &  ANS  &  11  &  6.984 $\pm$ 0.012  &  48  &  5.409 $\pm$ 0.036  \\
59439.812  &  ANS  &  11  &  7.405 $\pm$ 0.014  &  48  &  5.694 $\pm$ 0.019  \\
59439.881  &  TJO  &  11  &  7.420 $\pm$ 0.012  &   -   &   -                \\
59440.817  &  TJO  &   -   &          -         &  69  &  6.054 $\pm$ 0.014  \\
59440.818  &  ANS  &  12  &  7.676 $\pm$ 0.008  &  69  &  5.993 $\pm$ 0.011  \\
59440.840  &  ANS  &  12  &  7.566 $\pm$ 0.008  &  69  &  5.857 $\pm$ 0.026  \\
59441.817  &  TJO  &   -   &         -          &  87  &  6.268 $\pm$ 0.014  \\
59442.936  &  TJO  &  13  &  8.103 $\pm$ 0.012  &  91  &  6.407 $\pm$ 0.011  \\
59443.824  &  ANS  &  14  &  8.192 $\pm$ 0.008  &  89  &  6.434 $\pm$ 0.011  \\
59443.886  &  TJO  &  14  &  8.153 $\pm$ 0.012  &  89  &  6.429 $\pm$ 0.011  \\
59444.313  &  ANS  &  14  &  8.268 $\pm$ 0.008  &  89  &  6.486 $\pm$ 0.011  \\
59444.926  &  TJO  &  14  &  8.330 $\pm$ 0.013  &  83  &  6.538 $\pm$ 0.011  \\
\hline
\end{tabular}
\end{table}

% Extended data Table 8

%
\begin{table}
\caption{Log-book of spectroscopic observations, and 
expansion velocity of the H$_\alpha$, H$_\beta$ and He\,I\,5876$\lambda$  P-Cygni profiles. A conservative error of 250 km\,s$^{-1}$ has been associated to all velocities.}
\begin{tabular}{|l c c c|}
\hline 
\textbf{Telescope}  &    Serra la Nave  & Varese & \\
 \textbf{Spectrograph} &  CAOS & Echelle  & \\
 $\mathbf{R = \lambda/\Delta\lambda}$ &  45\,000 & 18\,000 &  \\
 \textbf{Range} & 400-900 nm& 425-890 nm& \\     \hline
    &\multicolumn{2}{c|}{MJD} &   Expansion velocity [km\,s$^{-1}$]        \\\hline
          &            &  59435.837  &  4250 \\
          &            &  59436.820  &  4600 \\
          &            &  59437.807  &  4750 \\
          &            &  59438.830  & 4000 \\
          &            &  59439.808  & 3000 \\
          &            &  59440.867  &  2800 \\
          &            &  59441.810  & 2700 \\
          & 59442.838  &  59442.824  &  2700 \\
          & 59443.821  &  59443.806  & 2700 \\
          & 59444.850  &  59444.810  &  2600 \\
          & 59445.852  &  59445.853  &  2500 \\
          & 59446.817  &  59446.808  &  2500 \\
          &            &  59447.801  &  2500 \\
          &            &  59448.796  &  2400 \\
          &            &  59449.794  &  2400 \\
          & 59450.822  &  59450.814  &  2400 \\
          &            &  59451.796  &  2400 \\
          &            &  59452.785  &  2400 \\
          & 59454.804  &             &  2300 \\
          & 59455.792  &             &  2300 \\
          & 59459.858  &             &  2100 \\
          & 59467.830  &             &  2100 \\
          & 59470.835  &             &  2000 \\
%
%
%\textbf{Telescope} &     TNG  &    Serra la Nave  & Varese & \\
% \textbf{Spectrograph} & HARPS-N &  CAOS & Echelle  & \\
% $\mathbf{R = \lambda/\Delta\lambda}$ & 115\,000 &  45\,000 & 18\,000 &  \\
% \textbf{Range} &370-690 nm & 400-900 nm& 425-890 nm& \\     \hline
%    &\multicolumn{3}{c|}{MJD} &   Expansion velocity [km\,s$^{-1}$]        \\\hline
%&          &            &  59435.837  &  4250 \\
%&          &            &  59436.820  &  4600 \\
% &         &            &  59437.807  &  4750 \\
% &         &            &  59438.830  & 4000 \\
% &         &            &  59439.808  & 3000 \\
%&59440.907 &            &  59440.867  &  2800 \\
%&          &            &  59441.810  & 2700 \\
%&          & 59442.838  &  59442.824  &  2700 \\
%&          & 59443.821  &  59443.806  & 2700 \\
%&          & 59444.850  &  59444.810  &  2600 \\
%&          & 59445.852  &  59445.853  &  2500 \\
%&          & 59446.817  &  59446.808  &  2500 \\
%&          &            &  59447.801  &  2500 \\
%&          &            &  59448.796  &  2400 \\
%&          &            &  59449.794  &  2400 \\
%&          & 59450.822  &  59450.814  &  2400 \\
%&          &            &  59451.796  &  2400 \\
%&          &            &  59452.785  &  2400 \\
%&          & 59454.804  &             &  2300 \\
%&          & 59455.792  &             &  2300 \\
%&          & 59459.858  &             &  2100 \\
%&          & 59467.830  &             &  2100 \\
%&          & 59470.835  &             &  2000 \\
%
\hline
\end{tabular}
\label{logbook_spectroscopy}
\end{table}

% Extended data Table 9

%

\begin{table}%[htbp]
\centering
\caption{Daily \fermi{} and MAGIC joint spectral fit results. Individual columns give normalization $f_0$ at normalization energy $E_0 = 130$\,GeV, slope $\alpha$ at $E_0$, curvature parameter $\beta$ and goodness of fit ($\chi^2/\mathrm{N_{dof}}$). Errors represent 1-sigma statistical uncertainties in the fits.}
\label{tab:daily_sed_fm}
\begin{tabular}{|c c c c r|}

\hline
MJD & $f_0$[$10^{-10}$TeV$^{-1}$cm$^{-2}$s$^{-1}$] & $\alpha$ &  $\beta$  & $\chi^2/\mathrm{N_{dof}}$\\  %&$E_0$ [GeV]
\hline
% 59435.94 - 59435.98 & $3.15\pm 0.79$ & $3.82\pm0.12$ & $0.658\pm0.026$& 6.21/5 \\ %p=0.29
% 59436.89 - 59437.04 & $2.62\pm 0.47$ & $3.71\pm0.10$ & $0.625\pm0.027$ & 17.46/5 \\ %p=0.0037
% 59437.89 - 59438.03 & $3.18\pm 0.52$ & $3.59\pm0.10$ & $0.614\pm0.029$ & 3.90/5\\ %p=0.56
% 59438.88 - 59439.02 & $2.98\pm 0.47$ & $3.39\pm0.12$ & $0.569\pm0.040$ & 11.62/5\\ %p=0.040
% updated 20210921
 59435.94 - 59435.98 & $5.4\pm 1.3$ & $3.86\pm0.13$   & $0.194\pm0.019$& 6.1/6 \\ % 6.15
 59436.89 - 59437.04 & $4.54\pm 0.78$ & $3.73\pm0.11$ & $0.175\pm0.020$ & 16.4/6 \\ %16.38
 59437.89 - 59438.03 & $5.37\pm 0.85$ & $3.64\pm0.12$ & $0.173\pm0.020$ & 3.7/6\\ %3.72
 59438.88 - 59439.02 & $5.00\pm 0.78$ & $3.44\pm0.14$ & $0.147\pm0.027$ & 10.8/6\\ %10.81
\hline
 59435.94 - 59439.02 & $5.08\pm0.45$ & $3.697\pm0.059$& $0.175\pm0.010$ & 9.3/6 \\ % 9.25/5
\hline
\end{tabular}
\end{table}

% Extended data Table 10

%
\begin{table}%[htbp]
\centering
\caption{Summary of the nova parameters used for the modeling of nova gamma-ray four day averaged spectrum (see Fig.~3). Parameters marked with an asterisk have modified values in the night-by-night modeling (see  EDF~\ref{figs:sed_daily})}
\label{tab:nova_par}    
\begin{tabular}{|c c c|}
\hline
Parameter & Symbol & Value \\
\hline
Distance & $d$ & $2.45\,$kpc \\
Photosphere radius & $R_{ph}$ & $200\,R_\odot$ \\
Photosphere temperature & $T_{ph}$ & 8460\,K* \\
time after nova explosion & $t$ & 3\,d* \\
Expansion velocity & $v_{sh}$ & $4500\,\mathrm{km\,s^{-1}}$ \\
Mass of nova ejecta & $M_{ej}$ & $10^{-6}\,M_\odot$ \\
Confinement factor & $h$ & 0.1 \\
\hline
\end{tabular}
\end{table}

% Extended data Table 11

%

\end{document}